\documentclass[a4paper,11pt,reqno]{article}
\usepackage{a4wide}

\usepackage{amsmath,amsfonts,amssymb}
\usepackage[T1]{fontenc}
\usepackage[p,osf]{cochineal}
\usepackage[varqu,varl,var0]{inconsolata}
\usepackage[scale=.95,type1]{cabin}
\usepackage[vvarbb]{newtxmath}
\usepackage[cal=boondoxo]{mathalfa}
\usepackage[mathscr]{euscript}
\usepackage{bbold} 

 \usepackage{soul}

\usepackage{lettrine} 
\usepackage{pict2e}

\makeatletter
\DeclareRobustCommand{\loplus}{\mathbin{\mathpalette\dog@lsemi{+}}}
\DeclareRobustCommand{\lotimes}{\mathbin{\mathpalette\dog@lsemi{\times}}}
\DeclareRobustCommand{\roplus}{\mathbin{\mathpalette\dog@rsemi{+}}}
\DeclareRobustCommand{\rotimes}{\mathbin{\mathpalette\dog@rsemi{\times}}}

\newcommand{\dog@rsemi}[2]{\dog@semi{#1}{#2}{-90,90}}
\newcommand{\dog@lsemi}[2]{\dog@semi{#1}{#2}{270,90}}
\newcommand{\dog@semi}[3]{%
  \begingroup
  \sbox\z@{$\m@th#1#2$}%
  \setlength{\unitlength}{\dimexpr\ht\z@+\dp\z@\relax}%
  \makebox[\wd\z@]{\raisebox{-\dp\z@}{%
    \begin{picture}(1,1)
    \linethickness{\variable@rule{#1}}
    \roundcap
    \put(0.5,0.5){\makebox(0,0){\raisebox{\dp\z@}{$\m@th#1#2$}}}
    \put(0.5,0.5){\arc[#3]{0.5}}
    \end{picture}%
  }}%
  \endgroup
}
\newcommand{\variable@rule}[1]{%
  \fontdimen8  
  \ifx#1\displaystyle\textfont3\else
    \ifx#1\textstyle\textfont3\else
      \ifx#1\scriptstyle\scriptfont3\else
        \scriptscriptfont3\relax
  \fi\fi\fi
}
\makeatother

\usepackage{nicefrac}
\usepackage{setspace}
\usepackage{datetime}
\usepackage[nosort]{cite}

\usepackage{upgreek}

\usepackage{comment}

\usepackage[euler]{textgreek}
\usepackage[breaklinks=true]{hyperref}

\numberwithin{equation}{section}

\hypersetup{
			colorlinks=true,
			urlcolor=blue,
			citecolor=magenta,
			linkcolor=blue,
     }

\let\oldsqrt\sqrt
\def\sqrt{\mathpalette\DHLhksqrt}
\def\DHLhksqrt#1#2{%
\setbox0=\hbox{$#1\oldsqrt{#2\,}$}\dimen0=\ht0
\advance\dimen0-0.2\ht0
\setbox2=\hbox{\vrule height\ht0 depth -\dimen0}%
{\box0\lower0.4pt\box2}}

\def\crbig{\\\noalign{\vspace{1.1mm}}}
\newcommand{\RNum}[1]{\uppercase\expandafter{\romannumeral #1\relax}}

\author{
  \begin{minipage}{.97\linewidth}
    \vspace{1cm}
       \begin{center}
      \begin{small}
             \textbf{Nehal Mittal},$^{1}$ 
     \textbf{P. Marios Petropoulos},$^2$ 
      \textbf{David Rivera-Betancour}$^2$ and 
      \textbf{Matthieu Vilatte}$^{2,3}$
              \end{small}
    \end{center}
    \vspace{0.5cm}
    \hspace{2.4cm}\begin{minipage}{.7\linewidth}
\begin{center}     {\it \begin{footnotesize}
\hbox{\kern-1.8cm\vbox{\vskip0cm
 \begin{itemize}
               \item[$^1$]Laboratoire Kastler Brossel -- LKB\\
               Coll\`ege de France, CNRS\footnote{\emph{Centre National de la Recherche Scientifique}, Unit\'e Mixte de Recherche UMR 8552.}\\
               ENS-PSL University, Sorbonne Universit\'e\\
               11 Place Marcelin Berthelot, 75005 Paris, France\\ 
                           \vskip0.25cm
      \end{itemize}}
\kern-3.2cm\vbox{
\begin{itemize}
 \item[$^2$]Centre de Physique Th\'eorique -- CPHT\\ 
        \'Ecole polytechnique, CNRS\footnote{\emph{Centre National de la Recherche Scientifique}, Unit\'e Mixte de Recherche UMR 7644.}\\
        Institut Polytechnique de Paris\\
        91120 Palaiseau Cedex, France         
      \end{itemize}
      \vskip0.cm
}}
     \end{footnotesize}}
\end{center}
    \end{minipage}
    \vspace{0.5cm}\begin{minipage}{.7\linewidth}
\begin{center}     
{\it \begin{footnotesize}
\hbox{\kern0.6cm\vbox{\vskip2cm
}
\kern3.9cm\vbox{
\begin{itemize}
 \item[$^3$]Division of Theoretical Physics\\School of Physics\\
  Aristotle University of Thessaloniki\\ 
  54124 Thessaloniki, Greece
      \end{itemize}\vskip0.05cm
}
}
     \end{footnotesize}}
\end{center}
     \end{minipage}
  \end{minipage}
}

\title{\vspace{1.5cm}
 \boldmath \begin{LARGE}
    \textbf{\textsc{Ehlers, Carroll, Charges and Dual Charges}}
  \end{LARGE} \unboldmath
}

\date{}

\begin{document}


\begin{titlepage}
\maketitle
\thispagestyle{empty}

 \vspace{-12.cm}
  \begin{flushright}
  CPHT-RR047.072022\\
  \end{flushright}
 \vspace{11.cm}

\begin{center}
\textsc{Abstract}\\  
\vspace{1. cm}	
\begin{minipage}{1.0\linewidth}

We unravel the boundary manifestation of Ehlers' hidden M\"obius symmetry present in four-dimensional Ricci-flat spacetimes that enjoy a time-like isometry and are Petrov-algebraic. This is achieved in a designated gauge, shaped in the spirit of flat holography, where the Carrollian three-dimensional nature of the null conformal boundary is manifest and covariantly implemented. The action of the M\"obius group is local on the space of Carrollian boundary data, among which the Carrollian Cotton tensor plays a predominent role.
The Carrollian and Weyl geometric tools introduced for shaping an appropriate gauge, as well as the boundary conformal group, which is $\text{BMS}_4$, allow to define electric/magnetic, leading/subleading towers of charges directly from the boundary Carrollian dynamics and explore their behaviour under the action of the M\"obius duality group.

\end{minipage}
\end{center}


\end{titlepage}

\onehalfspace

\begingroup
\hypersetup{linkcolor=black}
\tableofcontents
\endgroup
\noindent\rule{\textwidth}{0.6pt}

\section{Introduction}

Hidden symmetries have a long history in relativistic theories of gravity, which started with the seminal work of Ehlers in the late fifties \cite{Ehlers}. It was shown in this article that in the presence of an isometry, vacuum Einstein's equations were invariant under M\"obius transformations. This observation triggered an important activity in several directions. In line with the sixties' renaissance of general relativity, it opened the way for solution-generating techniques applicable to vacuum Einstein's equations  \cite{Geroch, Geroch:1972yt}. This was soon generalized to situations with more commuting Killing fields \cite{Ernst:1967wx,Ernst:1968}  -- and bigger hidden symmetry group, providing the system with remarkable and unexpected integrability properties \cite{belinskii, Maison:1978es, maison2, Mazur:1982,Breitenlohner:1986um,  Nicolai:1991tt, Bernard:2001pp, Alekseev:2010mx, Katsimpouri:2012ky}. 
The underlying deep origin for the above pattern was unravelled with the advent of higher-dimensional supergravity theories, and is rooted in the reduction mechanism. This has revealed a wide class of hidden groups, among which the exceptional play a prominent role (see e.g. \cite{Julia:1980gr,Julia:1982gx,Breitenlohner:1987dg}, or \cite{Bossard:2021ebg} for a more recent presentation and further references). 

The integrable sector of Einstein's equations is only a tiny fraction of their solution space. Unveiling the latter, in conjunction with its asymptotic symmetries and conserved charges, has been in the very early agenda of general relativity. It shares features with gauge theories because of general covariance, and has led Bondi to come out with his homonymous gauge, where a systematic resolution of Einstein's equations is possible as an expansion in powers of a radial coordinate.\footnote{Other canonical gauges are Fefferman--Graham or Newman--Unti --  see e.g. Refs. \cite{Ruzziconi:2019pzd, Ciambelli:2022vot} for a review and more complete reading suggestions on this subject.} This delivers a set of functions of time and angular coordinates, obeying first-order time-evolution equations. For Ricci-flat spacetimes (asymptotically locally flat) this set is infinite, but it is finite for Einstein spacetimes with negative cosmological constant (asymptotically locally anti-de Sitter). 

In modern language, the set of functions necessary for reconstructing the solution are said to be defined on the conformal boundary of the spacetime. In the asymptotically flat instance the conformal boundary is null infinity and features a Carrollian three-dimensional hypersurface.\footnote{The original observation that triggered this ``flat-holography'' activity is described in Refs. \cite{Bagchi, Bagchi2}. A more systematic analysis in four dimensions was presented in \cite{CMPPS2}, which set the foundations for a Carrollian description of the dual theory, and provides a more complete reference list. Up-do-date developments in this vein are Refs. \cite{DFHR1, Indiancarcel, DFHR2}.} Bulk Einstein dynamics is therefore traded for boundary effective Carrollian conformal field dynamics. This statement is accurate when discussing Einstein's equations. Whether it could be promoted to a holographic principle akin to the better known AdS/CFT involving asymptotically anti-de Sitter spacetimes and conformal field theories defined on their time-like conformal boundary is a timely subject, currently under scrutiny.  
 
How do hidden symmetries such as Ehlers' act on the Carrollian boundary data? This is the central question we would like to address in the present work. The conformal symmetries of the boundary reflect the asymptotic symmetries of the bulk. These define for instance the $\text{BMS}_{4}$ algebra (Bondi--van der Burg--Metzner--Sachs \cite{Bondi1962, Sachs1962-2, Sachs1962}), which is isomorphic to the conformal Carroll algebra in three dimensions $\mathfrak{ccarr}(3)$ (see \cite{Duval:2014uva, Duval:2014lpa }), and emerges upon appropriate fall-off conditions. From this perspective, wondering how the bulk hidden symmetries are embraced by the Carrollian boundary and what their interplay is with $\text{BMS}_{4}\equiv \mathfrak{ccarr}(3)$, is both natural and relevant. 

There is yet another motivation for pursuing  this analysis. Following Geroch \cite{Geroch, Geroch:1972yt}, the action of some Ehlers subgroup is a duality rotation in the plane of gravitational electric and gravitational magnetic charges, as are e.g. the mass and the nut charge. Ricci-flat spacetimes possess in fact multiple infinities of charges (not necessarily conserved), incarnated  in pairs of electric and magnetic representatives, and originating from the infinitely many independent ``subleading'' degrees of freedom  necessary for reconstructing the bulk solution, as well as  the infinitely many generators of the asymptotic symmetry group. This picture has been widely conveyed through the work of Godazgar--Godazgar--Pope \cite{Godazgar:2018vmm, Godazgar:2018qpq, Godazgar:2018dvh} (see also \cite{Kol:2019nkc,Godazgar:2020gqd,Godazgar:2020kqd,Oliveri:2020xls,Kol:2020vet}) and amply deserves to be reconsidered in the light of hidden symmetries. The remarkable fact is here that such an analysis can be conducted exclusively on the boundary, where the charges are constructed (see e.g. \cite{CM1}) using the boundary dynamics combined with the three-dimensional-boundary Carrollian conformal isometries, the latter being always generated by the infinite-dimensional algebra $\text{BMS}_{4}\equiv \mathfrak{so}(3,1) \loplus \text{supertranslations}$ 
 \cite{Ciambelli:2019lap}. Translating the Ehlers group on the null boundary forcedly exhibits a mapping among the infinite towers of charges, which is obscured in a bulk approach. The boundary Carrollian geometry provides the most suitable language for clarifying these properties.
  
In the present work, we will analyse along the above lines the integrable sector of Ricci-flat spacetimes possessing a time-like Killing field, whose congruence coincides with the boundary Carrollian fiber. This sector is obtained by setting conditions on the boundary data, which ultimately guarantee that the infinite series in powers of the radial coordinates is resummed. The boundary conditions involve the Carrollian boundary Cotton tensor and the Carrollian boundary momenta (see \cite{CMPPS2}), which both enter the boundary computation of the charges associated with the solution at hand. They
rephrase the special structure of bulk Weyl tensor\footnote{The interplay between the bulk Weyl tensor, expanded in powers of the radial coordinate, and the boundary Cotton plus energy--momentum tensors for Einstein spacetimes was disclosed in \cite{deHaro:2007fg,Mansi:2008br, Mansi:2008bs} -- see also \cite{Petropoulos:2014yaa}. There is no rigorous similar statement for the Ricci-flat instance since the Carrollian relatives of the Cotton tensor have not yet been thoroughly investigated.} and unsurprisingly lead to algebraic Ricci-flat spacetimes. 
Although this class leaves interesting cases aside, it captures the main feature of the Ehlers-group boundary manifestation. The latter turns out to be an \emph{algebraic} transformation mixing e.g. the Carrollian Cotton scalar and the Bondi mass aspect. 
This mixing is transmitted to other boundary observables, including the charges through their boundary expression, and completes the picture of the bulk-and-boundary action of the hidden group. The action of the M\"obius group, generically non-local on the four-dimensional Ricci-flat metric, is therefore \emph{local} on the boundary -- as it is on the three-dimensional sigma-model of the reduction along the bulk Killing congruence.

The starting point of our study is a reminder on the Ehlers group and the Geroch method, as they emerge in the reduction of Ricci-flat spacetimes along orbits of one-dimensional groups of motions.  We next move and describe the bulk-to-boundary relationship for four-dimensional Ricci-flat spacetimes. This requires the use of a gauge (we call it ``modified Newman--Unti'' or ``covariantized'') in which the three-dimensional Carrollian boundary geometry is ostensible. The bulk metric in this gauge is manifestly covariant with respect to the boundary Carrollian diffeomorphisms and to the boundary Weyl transformations. The Carrollian boundary dynamics induced by the bulk Einstein's equations is the following item in our agenda, which further enables us to define sets of charges and dual charges -- electric and magnetic. Finally, using the available tools for a Ricci-flat spacetime enjoying a time-like isometry, we translate the action of the bulk M\"obius transformations onto the boundary observables. This analysis is performed for the integrable sector (resummable metrics) and is based on a specific class of time-like Killing fields. 

Most of our investigation relies on rather unusual geometric tools, which have been developed recently in the framework of Carroll structures. We have sorted them out in a first appendix, valid for any dimension $d+1$. Carrollian dynamics and conservation properties, necessary for describing the boundary perspective as inherited from bulk Einstein's equations, is summarized in the second appendix. The third appendix is specific to three dimensions with emphasis on the Carrollian Cotton tensors.

\section{Ehlers and Geroch}\label{EG}

We remind here Geroch' generalization of Ehlers' work following   \cite{Geroch}. We consider a four-dimensional pseudo-Riemannian manifold $(\mathcal{M},\text{g})$ possessing an isometry generated by  a time-like\footnote{The described procedure goes through in the same fashion with space-like isometries, but keeping the two options would bring unnecessary multiplication of indices without shedding more light on our purpose.} Killing vector field $\upxi$. The latter has norm and twist -- here $A,B,\ldots\in \{0,\ldots,3\}$:
\begin{eqnarray}
\label{lambda}
\lambda&=&\xi^A\xi_A,\\
w_A&=&
\eta_{ABCD}\xi^B \nabla^C \xi^D,
\label{doublew}
\end{eqnarray}
where $\eta_{ABCD}=\sqrt{-g}\, \epsilon_{ABCD}$ ($\epsilon_{0123}=1$). Assuming the spacetime be Ricci-flat,\footnote{This property actually holds more generally for Einstein spacetimes \cite{Leigh:2014dja}.} one shows that the one-form $\text{w}=w_A\text{d}x^A$ is closed so locally exact, hence
\begin{equation}
\label{omega}
   \text{w} = \mathrm{d}\omega
\end{equation}
with $\omega$ a scalar function. 

We define the three-dimensional space $\mathcal{S}$ as the quotient $\nicefrac{\mathcal{M}}{\text{orb}(\upxi)}$. This coset space is not a subspace of $\mathcal{M}$ unless $\upxi$ is hypersurface-orthogonal, which would imply zero twist with $\mathcal{S}$ the orthogonal hypersurface. A natural metric on $\mathcal{S}$ is induced by $\text{g}$ of $\mathcal{M}$:
\begin{equation}\label{met-h}
h_{AB}=g_{AB}-\frac{\xi_A\xi_B}{\lambda},
\end{equation}
which defines the projector onto $\mathcal{S}$ as 
\begin{equation}\label{proj-h}
h^B_A=\delta^B_A-\frac{\xi^B\xi_A}{\lambda}.
\end{equation}
The fully antisymmetric tensor for \eqref{met-h} is  $  \eta_{ABC}= \frac{-1}{\sqrt{-\lambda}}\eta_{ABCD}\xi^D$.

Tensors of $\mathcal{M}$, transverse and invariant with respect to $\upxi$, are in one-to-one correspondence with tensors on $\mathcal{S}$. If $\text{T}$ is a tensor of  $\mathcal{S}$, the covariant derivative $\mathcal{D}$ defined following this correspondence, 
\begin{equation}\label{covd-h}
\mathcal{D}_C T_{A_1\ldots A_p}^{\hphantom{A_1\ldots A_p}B_1\ldots B_q} =
h_C^L h_{A_1}^{M_1}\dots  h_{A_p}^{M_p}h^{B_1}_{N_1}\dots  h^{B_q}_{N_q}
\nabla_L T_{M_1\ldots M_p}^{\hphantom{M_1\ldots M_p}N_1\ldots N_q} 
\end{equation}
with $\nabla$ the Levi--Civita connection on $(\mathcal{M},\text{g})$, coincides with the Levi--Civita connection on $(\mathcal{S},\text{h})$. This sets a relationship between the Riemann tensor on $\mathcal{S}$ and the Riemann tensor on $\mathcal{M}$, generalizing thereby the Gauss--Codazzi equations to the instance where $\upxi$ is not-hypersurface orthogonal:
\begin{equation}
\label{R-h}
\mathcal{R}_{ABCD}
= h_{[A}^{\hphantom{[}P}h_{B]}^{Q}
h_{[C}^{\hphantom{[}R}h_{D]}^{S}
\left(R_{PQRS}+\tfrac{2}{\lambda}\left(
\nabla_P\xi_Q
\nabla_R\xi_S
+\nabla_P\xi_R
\nabla_Q\xi_S
\right)\right)
\end{equation}
(the calligraphic  letters refer to curvature tensors of $\mathcal{S}$). 

The Ricci-flat dynamics for $g_{AB}$ is recast in the present framework in terms of\footnote{With our conventions, this metric is definite-negative.}
\begin{equation}
\label{met-h-tilde}
\tilde{h}_{AB}=\lambda h_{AB},
\end{equation}
as well as $\omega$ and $\lambda$ viewed as fields on $\mathcal{S}$, 
packaged in 
\begin{equation}
\label{tau}
\tau = \omega + \text{i} \lambda,
\end{equation}
and obeying the following equations:\footnote{Equations \eqref{eq: funda equa} can be reached by varying a three-dimensional sigma-model action defined on $\mathcal{S}$. This is at the heart of many developments about integrability and hidden symmetries -- see the already quoted literature for more information.}
\begin{equation}
\label{eq: funda equa}
\begin{array}{rcl}
\displaystyle{ \tilde{\mathcal{R}}_{AB} }&=&\displaystyle{ -\frac{2}{(\tau - \bar{\tau})^{2}}\tilde{\mathcal{D}}_{(A}\tau \tilde{\mathcal{D}}_{B)}\bar{\tau},}  \crbig
\displaystyle{ \tilde{\mathcal{D}}^2\tau }&=&\displaystyle{\frac2{\tau - \bar{\tau}}\tilde{\mathcal{D}}_{M}\tau \tilde{\mathcal{D}}_{N} \tau\tilde{h}^{MN}} .
\end{array}
\end{equation}
The first results from \eqref{R-h}, while the second is obtained by a direct computation of the $\mathcal{S}$-Laplacian acting on $\tau $. Here
$\tilde{\mathcal{D}}_A$ and $\tilde{\mathcal{R}}_{AB}$ are the Levi--Civita covariant derivative and the Ricci tensor associated with the metric $\tilde{h}_{AB}$ displayed in \eqref{met-h-tilde}.
 
Equations \eqref{eq: funda equa} feature two important properties. The first, due to Ehlers \cite{Ehlers}, is the invariance under transformations maintaining $\tilde{h}_{AB}$ \emph{unaltered} and mapping $\tau $ into
\begin{equation}
\tau'=\frac{\alpha\tau +\beta}{\gamma\tau+\delta}\, ,\quad
\begin{pmatrix}      \alpha & \beta \\ \gamma & \delta  \end{pmatrix} \in SL(2,\mathbb{R}). 
\label{ehlers}
\end{equation}
This is the original instance where a  \emph{hidden} group, $SL(2,\mathbb{R})$, reveals upon reduction with respect to an isometry. 
The second, described by Geroch in \cite{Geroch, Geroch:1972yt},  is the method for reversing the reduction process, and finding a Ricci-flat four-dimensional spacetime with an isometry, starting from any solution of Eqs.~\eqref{eq: funda equa} encoded in   $\omega' + \text{i} \lambda' = \tau'$ and $h_{AB}^{\prime}=\frac{1}{\lambda'} \tilde{h}_{AB}$. To this end, one shows that the $\mathcal{S}$-two-form defined as
\begin{equation}\label{F}
F_{AB}^\prime=\frac{1}{(-\lambda')^{\nicefrac{3}{2}}} \eta_{ABC}^\prime \mathcal{D}^C \omega'
 \end{equation}
is \emph{closed}. Thus, locally
\begin{equation}\label{eta}
\text{F}'=\text{d}\upeta'.
\end{equation}
The one-form field $\upeta'$, defined on $\mathcal{S}$, can be promoted to a field on $\mathcal{M}$ by adding the necessary exact piece such that its normalization be
\begin{equation}\label{eta-norm}
\xi^A \eta_A^\prime=1.
\end{equation}
This defines a new Killing field on $\mathcal{M}$
\begin{equation}\label{newkil}
\upxi'=  \lambda' \upeta'
\end{equation}
and the new four-dimensional metric reads:\footnote{The consequence of M\"obius transformations on the Weyl tensor has been investigated in Ref. \cite{Mars}.}
\begin{equation}\label{newmet}
g_{AB}'=h_{AB}'+\frac{\xi_A'\xi_B'}{\lambda'}.
\end{equation}

Closing this executive reminder, we would like to add a remark. The $SL(2,\mathbb{R})$ is hidden from the four-dimensional perspective, but explicit in the three-dimensional sigma-model, materialized here in Eqs.   \eqref{eq: funda equa}. Nevertheless, part of this group is in fact visible in four dimensions because it acts as four-dimensional diffeomorphisms; part is creating genuinely different Ricci-flat solutions. This can be illustrated in the concrete example of Schwarzschild--Taub--NUT solutions with mass $M$ and nut charge $n$. The compact  
subgroup  of rotations $\left(\begin{smallmatrix}      \cos \chi & \sin \chi \\ -\sin \chi & \cos \chi  \end{smallmatrix}\right) \in SO(2)\subset SL(2,\mathbb{R})$ induces rotations of angle $2\chi$ in the parameter space $(M,n)$, while  non-compact transformations $\left(\begin{smallmatrix}      \alpha & \beta \\ 0 & \nicefrac{1}{\alpha}  \end{smallmatrix}\right)\in N \subset SL(2,\mathbb{R})$ act homothetically,  $(M,n)\to (\nicefrac{M}{\alpha} , \nicefrac{n}{\alpha} )$.

\section{Ricci-flat spacetimes and Carrollian dynamics}\label{RFCB}

\subsection{Bulk reconstruction and resummable Ricci-flat metrics}\label{brrm}

\subsubsection*{Choosing a covariant gauge}

Four-dimensional Ricci-flat metrics are generally obtained as expansions in powers of a radial coordinate, in a designated gauge, usually Bondi or Newman--Unti. Appropriate fall-offs are assumed, and the solution is expressed in terms of an infinite set of functions of time and angles, obeying some evolution equations, mirroring Einstein's equations (see \cite{Ruzziconi:2019pzd} for details and further references). Can one define a three-dimensional boundary, and describe covariantly this set of functions and their dynamics?

The answer to this question has been known to be positive for a long time in the case of Einstein spacetimes. It is best formulated in the Fefferman--Graham gauge \cite{PMP-FG1, PMP-FG2} -- see also \cite{ciamlei} for a Weyl-covariant extension of this gauge. The (conformal) boundary is a three-dimensional pseudo-Riemannian spacetime, and every order in the expansion brings a tensorial object with respect to the boundary geometry. All these are expressed in terms of two independent tensors: the first and second fundamental forms of the boundary, namely the boundary metric and the boundary energy--momentum tensor, which is covariantly conserved with respect to the associated Levi--Civita connection. This conservation translates those of Einstein's equations that have not been used in the process of taming the expansion.

The boundary covariance of the Fefferman--Graham gauge makes it elegant and suitable for holographic applications in the framework of anti-de Sitter/conformal-field-theory correspondence. Setting up a gauge that is covariant with respect to the boundary is therefore desirable as part of the effort to unravel a similar duality for asymptotically flat spacetimes.
In this case, the conformal boundary is at null infinity and is endowed with a Carrollian geometry \cite{Bagchi, Bagchi2}. 

Carroll structures \cite{Duval:2014uva, Duval:2014lpa, Ciambelli:2019lap, Duval:2014uoa, Bekaert:2014bwa,Bekaert:2015xua, Hartong:2015xda, Morand:2018tke, Herfray:2021qmp} consist of a $d+1$-dimensional manifold $\mathscr{M}= \mathbb{R} \times \mathscr{S}$  equipped with a degenerate metric. The kernel of the metric is  a vector field called  \emph{field of observers}. We will adopt coordinates $( t, \mathbf{x})$ and a metric of the form
\begin{equation}   
\label{cardegmet}
\text{d}\ell^2=a_{ij}( t, \mathbf{x}) \text{d}x^i \text{d}x^j,\quad i,j\ldots \in \{1,\ldots,d\}
\end{equation}
with kernel
\begin{equation}   
\label{kert}
\upupsilon
= \frac{1}{\Omega}\partial_t.
\end{equation} 
The coordinate system at hand is adapted to the space/time splitting. It is thus respected by Carrollian diffeomorphisms 
 \begin{equation}
\label{cardifs} 
t'=t'(t,\mathbf{x})\quad \text{and} \quad \mathbf{x}^{\prime}=\mathbf{x}^{\prime}(\mathbf{x})
\end{equation}
with Jacobian 
\begin{equation}
 \label{carj}
J(t,\mathbf{x})=\frac{\partial t'}{\partial t},\quad j_i(t,\mathbf{x}) = \frac{\partial  t'}{\partial x^{i}},\quad 
J^i_j(\mathbf{x}) = \frac{\partial x^{i\prime}}{\partial x^{j}}.
\end{equation}
The  \emph{clock form} is dual to the field of observers with $\upmu(\upupsilon)=-1$:
\begin{equation}   
\label{kertdual}
\upmu=-\Omega \text{d}t +b_i \text{d}x^i
\end{equation}
($\Omega$ and $b_i$ depend on $t$ and $\mathbf{x}$) and  incorporates an \emph{Ehresmann connection}, which is the background gauge field  
$\pmb{b}=b_i \text{d}x^i$.\footnote{A Carroll structure endowed with metric \eqref{cardegmet} and clock form \eqref{kertdual} is naturally reached in the Carrollian limit ($c\to 0$) of a pseudo-Riemannian spacetime  $\mathscr{M}$ in Papapetrou--Randers gauge 
$
\text{d}s^2 =- c^2\left(\Omega \text{d}t-b_i \text{d}x^i
\right)^2+a_{ij} \text{d}x^i \text{d}x^j
$,
where all functions are $x$-dependent with $x\equiv(x^0=ct,\mathbf{x})$. 
It should be noticed here that the degenerate metric could  generally have components along $\text{d}t$, which would in turn give $\partial_i$ components to the field of observers. In this instance, the above  Carrollian diffeomorphisms \eqref{cardifs} play no privileged role, and plain general covariance is at work -- without affecting the dynamics presented in App. \ref{conscar}. This option is sometimes chosen (see e.g. \cite{Hartong:2015xda} for a general approach, or \cite{CMPR, CMPRpos, Campoleoni:2022wmf} for an application  to three-dimensional Minkowski spacetime), but it is always possible to single out the time direction supported by the fiber of the Carrollian structure, i.e. distinguish time and spatial sections with no conflict with general covariance.\label{relpar}}
Carrollian tensors depend on time $t$ and space $\mathbf{x}$. They carry indices $i, j, \ldots $  lowered and raised with $a_{ij}$ and its inverse $a^{ij}$, and transform covariantly under \eqref{cardifs} with  $J_i^j$ and $J^{-1i}_{\hphantom{-1}j}$ defined in \eqref{carj}. The basics on Carrollian tensors and Carrollian covariant derivatives are summarized in App. \ref{carman}. In the following we will focus on $d=2$, corresponding to the three-dimensional conformal null boundary of a four-dimensional asymptotically flat spacetime, and further information on this instance is available in App. \ref{carcot3}. The boundary Carrollian covariance is part of the bulk general covariance, as inherited in the boundary geometry.

Fefferman--Graham gauge is only valid for Einstein spacetimes, on the one hand. On the other hand, Bondi and Newman--Unti gauges, applicable to Ricci-flat spacetimes, are not covariant with respect to the boundary, because the spatial section of the three-dimensional null boundary is locked. 
An alternative, still of the Eddington--Finkelstein type i.e. with a light-like radial direction, was introduced in the framework of fluid/gravity correspondence \cite{Haack:2008cp,Bhattacharyya:2008jc}, and made more systematic in the subsequent works
both in AdS \cite{Caldarelli:2012cm, Mukhopadhyay:2013gja, Gath:2015nxa, Petropoulos:2015fba} and for Ricci-flat spacetimes \cite{CMPPS2}. It is a sort of modified and slightly incomplete Newman--Unti gauge  \cite{CMPR, CMPRpos, Campoleoni:2018ltl} (see also \cite{Nguyen:2020hot,Geiller:2022vto} for other extensions of the Bondi or Newman--Unti  gauges). The time coordinate $t$ is actually a retarded time (usually spelled $u$) and coincides at the boundary with the Carrollian time used in \eqref{cardegmet},  \eqref{kert} and  \eqref{kertdual}.

We can summarize as follows the structure of the four-dimensional Ricci-flat solutions in the advertised gauge, up to order $\nicefrac{1}{r^2}$ ($G$ is four-dimensional Newton's constant):
 \begin{eqnarray}
\text{d}s^2_{\text{Ricci-flat}}\!\!&=&\upmu\left[2\text{d}r-\left(r\theta+\hat{\mathscr{K}}\right)\upmu+ \left(2r\varphi_i-2\ast\!\hat{\mathscr{D}}_{i}\ast\!\varpi-\hat{\mathscr{D}}_{j}\mathscr{C}^j_{\hphantom{j}i}\right)\text{d}x^i\right]\nonumber\\
&&+\mathscr{C}_{ij}\left(r\text{d}x^i\text{d}x^j-\ast\varpi\ast\!\text{d}x^i\text{d}x^j\right)+\left(r^2+\ast\varpi^2+\frac{\mathscr{C}_{kl}\mathscr{C}^{kl}}{8}\right)\text{d}\ell^2 \nonumber\\
&&+\frac{1}{r}\left[8\pi G\varepsilon\upmu^2+\frac{32\pi G}{3}\left(\pi_{i}-\frac{1}{8\pi G}\ast\! \psi_{i}\right)\text{d}x^i \upmu-\frac{16\pi G}{3}E_{ij}\text{d}x^i\text{d}x^j\right] 
\nonumber\\
&&
+ 
\frac{1}{r^2}\left(\ast \varpi c \upmu^2 +\cdots\right)+
\text{ O}\left(\frac{1}{r^3}\right),
\label{exp-ric-fl}
\end{eqnarray}
where the star designates a $d=2$ Carrollian Hodge duality as defined in Eq. \eqref{hodgeast}.\footnote{Referring to the complex coordinates introduced in  App. \ref{carcot3}, we chose the orientation as inherited from the parent Riemannian spacetime:  $\eta_{0\zeta\bar\zeta}=\Omega\sqrt{a}\epsilon_{0\zeta\bar\zeta}=\frac{\text{i}\Omega}{P^2}$, where $x^0=kt$.
\label{orient}}
As anticipated, this expression is neither in Bondi gauge (no determinant condition --  see \cite{Bondi1962,Sachs1962-2}), nor in Newman--Unti  ($g_{rt}=-\Omega\neq -1$ and $g_{ri}=b_i\neq0$, obtained using \eqref{kertdual} -- see \cite{NU1962}).\footnote{In all quoted Eddington--Finkelstein type of gauges, $\partial_r$ is tangent to a null geodesic congruence. In Newman--Unti and in modified Newman--Unti this congruence is affinely parameterized, in contrast to Bondi. In modified  Newman--Unti gauge,  
as opposed to the others, $\partial_r$ \emph{is not} hypersurface-orthogonal. Indeed, the metric-dual form to $\partial_r$ is $\upmu$, which has a twist because of $\Omega$ and $b_i$, the defining features of the gauge at hand: $\upmu\wedge \text{d}\upmu=\ast\varpi \eta_{ij}\text{d}x^i\wedge\text{d}x^j\wedge\upmu$ (we have used Eqs.  \eqref{dualcarconcomderf} and\eqref{scalar}).}
Delving into the details of this gauge would bring us outside the main purpose of the present work. We will rather explain the various ingredients appearing in the above expression and insist on their Carrollian-covariant nature.  This includes the account of the required boundary data and the description of the evolution equations they obey so that the bulk metric be Ricci-flat.

All quantities entering expression \eqref{exp-ric-fl} are defined on the conformal boundary and can be sorted as follows (see also the appendices for further information).
\begin{description}
\item[Carrollian geometry] The conformal boundary itself is part of the solution space. It is materialized in $a_{ij}$, $b_i$ and $\Omega$, accompanied with all attributes such as Carrollian connections and curvature tensors, Carrollian Cotton descendants   etc. -- see Apps. \ref{carman} and \ref{carcot3}. These are free data, without evolution equations, except for the restriction of \emph{vanishing Carrollian geometric shear as a consequence of Einstein's equations: $\xi_{ij}=0$}.\footnote{In BMS gauge, one would set $b_i=0$, $\Omega=1$, and $a_{ij}$ the round sphere.}
\item[Shear] The \emph{dynamic shear }is a symmetric and traceless Carrollian boundary tensor $\mathscr{C}_{ij}(t,\mathbf{x})$ not to be confused with the geometric shear $\xi_{ij}(t,\mathbf{x})$.\footnote{In Einstein spacetimes these two shears are proportional with the cosmological constant as a factor. In the asymptotically flat limit, the geometric shear is required to vanish, while the dynamic shear decouples.} It is a boundary emanation of the bulk $\partial_r$-congruence shear, and is completely free, although it sources the evolution equations of other tensorial data.  The dynamic shear carries information on the bulk gravitational radiation through the symmetric and traceless  \emph{Bondi-like news:} 
\begin{equation}
\label{news-rel-car}
\hat{\mathscr{N}}_{ij}=\frac{1}{\Omega}\hat{\mathscr{D}}_t \mathscr{C}_{ij}. 
\end{equation}
With these definitions, the shear and the news are supported by genuine boundary conformal Carrollian-covariant tensors (weight $-1$ and $0$), hence meeting the advertised expectations.\footnote{Notice that they do not exactly coincide with the original shear and news  defined in BMS gauge. They vanish in Robinson--Trautmann spacetimes expressed in the gauge at hand, which is their defining gauge, although these solutions are radiating.\label{RT-foot}} 

\item[Carrollian fluid] The boundary Carrollian fluid of Ricci-flat spacetimes is the descendant of the relativistic boundary fluid in Einstein spacetimes in the vanishing speed of light limit, supported by the conserved energy--momentum tensor $T_{\mu\nu}$. It is described  in terms of the energy density $\varepsilon$, the heat currents $Q^i$ and $\pi^i$, and the symmetric and traceless stress tensors $\Sigma^{ij}$ and $\Xi^{ij}$ \cite{CMPPS1, BigFluid}. The associated momenta of the fluid dynamics in the sense of App. \ref{conscar} are as follows:
\begin{equation}
\label{mom-fluid}
\Pi=\varepsilon,\quad \Pi^i=Q^i,\quad P^i=\pi^i,\quad \tilde\Pi^{ij}=-\Sigma^{ij},\quad \Pi^{ij}=\frac{\varepsilon}{2}a^{ij}-\Xi^{ij}.
\end{equation}
As opposed to the relativistic boundary fluid, however, the Carrollian fluid is not free, but sourced by the shear, the news and the  Carrollian Cotton descendants. Put differently, its dynamical equations are  \eqref{carEbiscon}, \eqref{carFcon}, \eqref{carGcon} and \eqref{carHcon} (at zero $\xi^{ij}$) with a non-vanishing right-hand side. These equations translate part of Einstein's, which furthermore impose\footnote{The presence of a non-vanishing energy flux $ \Pi^i=Q^i$ betrays the breaking of local Carroll boost invariance (see App. \ref{conscar}, footnote \ref{bbreaking}) in the boundary Carrollian dynamics associated with Ricci-flat spacetimes. This breaking accounts for bulk gravitational radiation, which in the boundary-covariant gauge designed here does not originate solely in the news \eqref{news-rel-car} but is also encoded in the Carrollian energy flux $\Pi^i=Q^i=\frac{1}{8\pi G} \ast \! \chi^{i}$ and the Carrollian stress $ \tilde\Pi^{ij}=-\Sigma^{ij}=-\frac{1}{8\pi G} \ast \! X^{ij} $ obeying Eq. \eqref{carHcon} or equivalently \eqref{carHcot}. In Robinson--Trautman spacetimes e.g., the gravitational radiation is exclusively rooted in the latter Cotton descendants -- see footnote \ref{RT-foot} and Ref. \cite{CMPPS2}.}
\begin{equation}
\label{visc-cargen-resum-expli}
Q^{i} =\frac{1}{8\pi G} \ast \! \chi^{i},\quad \Sigma^{ij} =\frac{1}{8\pi G} \ast \! X^{ij}  \quad   \Xi^{ij}=\frac{1}{8\pi G}\ast  \!  \Psi^{ij}.
\end{equation}
Three of the Carrollian fluid data are thus tuned in terms of the boundary geometry through the Carrollian Cotton descendants displayed in Eqs. \eqref{chi-f-Carrol}, \eqref{X-2-Carrol} and \eqref{Psi-2-Carrol}. Only two momenta remain independent ($\Pi=\varepsilon$ and $P^i=\pi^i$) and subject to two Carrollian-fluid evolution equations (\eqref{carEbiscon} and \eqref{carGcon}  with zero $\xi^{ij}$ and external force,\footnote{\label{flbal}We display for completeness these Carrollian equations, which coincide with Eqs. (2.53) and (2.50) of Ref. \cite{Compere:2019bua}, once translated from our gauge into the BMS gauge:
\begin{eqnarray}
 \frac{1}{\Omega}\hat{\mathscr{D}}_t\Pi
+\hat{\mathscr{D}}_i \Pi^{i}
&=& \frac{1}{16\pi G} \left(
\hat{\mathscr{D}}_i  \hat{\mathscr{D}}_j \hat{\mathscr{N}}^{ij}+ \mathscr{C}^{ij}
\hat{\mathscr{D}}_i \hat{\mathscr{R}}_j 
+\frac{1}{2} \mathscr{C}_{ij} \frac{1}{\Omega}\hat{\mathscr{D}}_t  \hat{\mathscr{N}}^{ij}
\right),
  \nonumber
  \\
\hat{\mathscr{D}}_j \Pi^{ij}+\frac{1}{\Omega}\hat{\mathscr{D}}_t P^i+2\ast\! \varpi \ast \!\Pi^{i}&=&\frac{1}{16\pi G}\bigg[\mathscr{C}^{ij}\hat{\mathscr{D}}_j\hat{\mathscr{K}}
+\ast\mathscr{C}^{ij}\hat{\mathscr{D}}_j\hat{\mathscr{A}}
-4\ast\!\varpi \ast\!\mathscr{C}^{ij}\hat{\mathscr{R}}_j
-\frac{1}{2}\hat{\mathscr{D}}^j\left(\hat{\mathscr{D}}_j\hat{\mathscr{D}}_k \mathscr{C}^{ik}
-\hat{\mathscr{D}}^i\hat{\mathscr{D}}^k \mathscr{C}_{jk}
\right)
 \nonumber \\
&&\hphantom{\frac{1}{16\pi G}\bigg[}+\mathscr{C}^{ij}\hat{\mathscr{D}}^k\hat{\mathscr{N}}_{jk}+\frac{1}{2}\hat{\mathscr{D}}^j\left( \mathscr{C}^{ik}\hat{\mathscr{N}}_{jk}\right)-\frac{1}{4}\hat{\mathscr{D}}^i\left( \mathscr{C}^{jk}\hat{\mathscr{N}}_{jk}\right)\bigg]
  \nonumber
 \end{eqnarray}
with $\Pi$, $\Pi^{i}$, $\Pi^{ij}$, $P^i$ as in Eqs. \eqref{mom-fluid} and \eqref{visc-cargen-resum-expli}.} often referred to as \emph{flux-balance equations}) out of the four -- the other two are automatically satisfied owing to the Cotton equations \eqref{carEbiscon}, \eqref{carFcon}, \eqref{carGcon} and \eqref{carHcon}  with  \eqref{mom-Cot}. These data are related to the Bondi mass and angular momentum aspects, $M(t,\mathbf{x})$ and $N^i(t,\mathbf{x})$:
\begin{eqnarray}
\label{maspect}
8\pi G \varepsilon&=&2M +\frac{1}{4}\mathscr{C}^{jk}\hat{\mathscr{N}}_{jk},
\\
8\pi G \pi^i& =&\ast\psi^i-N^i
\label{angmomaspect}
\end{eqnarray}
with $\psi^i$ given in \eqref{psi-f-Carrol}.

Similarly to the expansion of  Einstein spacetimes (in Fefferman--Graham or in the present gauge), fluid-related tensors appear \emph{at every order} and not  exclusively for $\nicefrac{1}{r}$, as expression \eqref{exp-ric-fl} might suggest.

\item[Further degrees of freedom] Contrary to the asymptotically anti-de Sitter case, the above fluid data \emph{are not} the only degrees of freedom besides the boundary geometry. An infinite number of Carrollian tensors are necessary to all orders in the radial expansion, as $E_{ij}(t,\mathbf{x})$ in \eqref{exp-ric-fl} at order $\nicefrac{1}{r}$, which obey Carrollian evolution -- flux-balance -- equations similar to those already displayed in footnote \ref{flbal}. These are dubbed ``Chthonian'' degrees of freedom.
\end{description}

We will not elaborate any further on the features of the expansion and the structure of the various evolution equations. The  covariantization with respect to boundary Carroll diffeomorphisms and Weyl covariance is a powerful tool,\footnote{The expression \eqref{maspect} matches with Eq. (42) of Ref. \cite{Freidel:2021qpz}, reached through a completely different logical path. Similarly the Carroll Cotton scalar $c$ given in \eqref{c-Carrol} plays here the role of the dual mass aspect, captured in (53) of the quoted reference.}  rooted in the bulk general covariance. It can be supplemented with the boundary-fluid hydrodynamic-frame invariance at the expense of giving up radically the complete bulk gauge fixing. This requires a modified and \emph{incomplete} Newman--Unti gauge, and has been performed for three bulk dimensions in Refs. \cite{CMPR, CMPRpos, Campoleoni:2022wmf, Campoleoni:2018ltl}. 

\subsubsection*{Resumming the series expansion}

In certain circumstances the series \eqref{exp-ric-fl} can be resummed. As advertised in the introduction, this occurs when  conditions are imposed on the boundary data, which enforce specific features for the bulk Weyl tensor:
\begin{enumerate}
\item the dynamic shear $\mathscr{C}_{ij}(t,\mathbf{x})$ should vanish, implying in particular the relation $M=4\pi G \varepsilon$;\label{bondmas}
\item all non-Carrollian-fluid related degrees of freedom should be discarded, as e.g. $E_{ij}(t,\mathbf{x})$; 
\item $N^i$ in \eqref{angmomaspect} should be set to zero, which amounts to demanding the Carrollian momentum $P^i=\pi^i$ be tuned with respect to a Carrollian Cotton descendant:\footnote{Although Eq. \eqref{pi-fluid-resum}, which secretly  tunes the bulk Weyl tensor, bares some resemblance with a self-duality condition, it isn't as the Ricci-flat spacetimes at hand are Lorentzian rather than Euclidean and this option is not available. \label{selfduality}}
\begin{equation}
\label{pi-fluid-resum}
 \pi^i =\frac{1}{8\pi G }\ast\psi^i.
 \end{equation}
\end{enumerate}

In the configuration reached with the above conditions, the remaining degrees of freedom are those describing the boundary Carrollian geometry (metric, field of observers and Ehresmann connection), and the Carrollian-fluid energy density i.e. the Bondi mass aspect. Expression 
\eqref{exp-ric-fl} is now resummed into an exact Ricci-flat spacetime of algebraically special type:\footnote{Details and examples are available in \cite{CMPPS2}. } 
 \begin{equation}
\text{d}s^2_{\text{res. Ricci-flat}}=\upmu\left[2\text{d}r+2\left(r\varphi_j-\ast\hat{\mathscr{D}}_{j}\ast\!\varpi\right)\text{d}x^j-\left(r\theta+\hat{\mathscr{K}}\right)\upmu\right]+\rho^2\text{d}\ell^2+ 
\frac{\upmu^2 }{\rho^2}\left[8\pi G\varepsilon r+\ast \varpi c\right]
\label{exp-ric-fl-res}
\end{equation}
with
\begin{equation}
\label{rho2car}
 \rho^2= r^2 +\ast \varpi^2.
\end{equation}  
Ricci flatness is guaranteed by the Carrollian fluid equations, which are now genuine conservation equations without forcing term \eqref{carEbiscon}, \eqref{carFcon}, \eqref{carGcon} and \eqref{carHcon}, where the momenta are  (using \eqref{visc-cargen-resum-expli} and \eqref{pi-fluid-resum})
\begin{equation}
\label{mom-fluid-res}
\Pi=\varepsilon,\quad \Pi^i=\frac{1}{8\pi G} \ast \! \chi^{i},\quad P^i=\frac{1}{8\pi G }\ast\psi^i,\quad \tilde\Pi^{ij}=-\frac{1}{8\pi G} \ast \! X^{ij},\quad \Pi^{ij}=\frac{\varepsilon}{2}a^{ij}-\frac{1}{8\pi G}\ast  \!  \Psi^{ij}.
\end{equation}
The same equations are identically obeyed by the Carrollian Cotton tensors \eqref{mom-Cot} and the geometric shear is vanishing. We are therefore left with two independent  equations, which are   \eqref{carEbiscon} and  \eqref{carGcon}:
\begin{eqnarray}
 \frac{1}{\Omega}\hat{\mathscr{D}}_t\varepsilon
+\frac{1}{8\pi G} \hat{\mathscr{D}}_i  \ast \! \chi^{i}
&=& 0 ,
  \label{carEbiscon-res} 
  \\
\hat{\mathscr{D}}_j \varepsilon-\frac{1}{8\pi G} \ast\!\hat{\mathscr{D}}_j c
&=&0 ,
  \label{carGconres} 
  \end{eqnarray}
where $c$ and $\chi^i$ are given in geometric terms in \eqref{c-Carrol} and \eqref{chi-f-Carrol}, and $\varepsilon$ is proportional to the Bondi-mass aspect, as stressed in item \ref{bondmas} above. Equations \eqref{carEbiscon-res} and \eqref{carGconres} are those displayed in footnote \ref{flbal} with vanishing right-hand side. 

From the above Eqs.  \eqref{carEbiscon-res} and \eqref{carGconres} as well as Eq.
\eqref{carEcot} one can foresee that the energy density $\varepsilon$ and the Carrollian Cotton scalar $c$ play dual roles. This will be formulated concretely  in Sec. \ref{RFCB} with reference to the boundary action of the M\"obius group. Anticipating this argument, we introduce the following Carrollian complex scalar $\hat \tau(t,\mathbf{x})$ and vector $\hat \chi^j(t,\mathbf{x})$:
\begin{eqnarray}
\label{taudef}
\hat \tau&=&-c+8\pi \text{i} G \varepsilon,\\
\hat \chi^j&=&\chi^j-\text{i} \ast\!\chi^j.
\label{chidef}
 \end{eqnarray}
The aforementioned equations  are thus recast as\footnote{The first of Eqs. \eqref{divtau} is flux-balance, driven exclusively by the Cotton vector $\hat \chi^j$ displayed in \eqref{chidef}. The loss phenomenon concerns both the mass aspect $\varepsilon$ and the ``magnetic-mass aspect'' $c$,  as captured in Eqs. (76) and (80) of \cite{Freidel:2021qpz} -- see also App. D of \cite{Barnich:2019vzx}.
As opposed to $\varepsilon$, the time evolution \eqref{carEcot} of the magnetic-mass aspect is not altered by $\mathscr{C}_{ij}$ and $\hat{\mathscr{N}}_{ij}$, in line with \cite{Ashtekar82}.}
\begin{equation}
\label{divtau}
\begin{array}{rcl}
& \frac{1}{\Omega}\hat{\mathscr{D}}_t \hat \tau =\hat{\mathscr{D}}_j\hat \chi^j,\qquad
\hat{\mathscr{D}}_j \hat \tau \hat{\mathscr{D}}^j  \hat \tau =0,&\\
&\hat{\mathscr{D}}_j \hat \tau \hat{\mathscr{D}}^j  \hat{\bar\tau} =
8\left(2\ast\!\varpi\ast\!\chi_i+ \frac{1}{\Omega}\hat{\mathscr{D}}_t \psi_i -\hat{\mathscr{D}}^j \Psi_{ji}
\right)
\left(2\ast\!\varpi\ast\!\chi^i+ \frac{1}{\Omega}\hat{\mathscr{D}}_t \psi^i -\hat{\mathscr{D}}_k \Psi^{ki}\right)
.&
 \end{array}
\end{equation}  
Acting with a second spatial derivative on \eqref{carGconres}  and using  \eqref{CWcontor}, we finally obtain 
\begin{equation}
\label{lapltau}
\hat{\mathscr{D}}^j \hat{\mathscr{D}}_j \hat \tau=2 \text{i} 
\left(
 \frac{1}{\Omega}\hat{\mathscr{D}}_t\ast\!\varpi \hat \tau -\hat{\mathscr{A}}\hat\tau
\right).
\end{equation}  
Let us mention for completeness that Eqs.  \eqref{carEbiscon-res} and \eqref{carGconres} coincide with Eqs. (29.16) and (29.15)
of \cite{Stephani:624239}.\footnote{For that purpose, the following identifications are necessary (in complex coordinates, as in App. \ref{carcot3}): $b_\zeta=-L$, $\ast \! \varpi=-\Sigma$, $\hat \tau =2(M +\text{i}m)$, $\Omega = 1$, $t=u$, whereas their radial coordinate is $\tilde r = r-r_0$ with $r_0(t, \zeta, \bar \zeta)$ the origin in the affine parameter of the geodesic congruence tangent to $\partial_r$.
\label{origin}} It is remarkable that complicated equations as the latter can actually be tamed into a simple  fluid conservation supplemented with a kind of self-duality requirement. It would have been unthinkable to reach such a conclusion without the null boundary analysis performed here and the corresponding Carrollian geometric tools. The latter provide definitely the natural language for unravelling asymptotically flat spacetimes.

A last comment before closing this section concerns the algebraic-special nature of the metric \eqref{exp-ric-fl-res}. This is proven thanks to the Goldberg--Sachs theorem using the null, geodesic and, in the resummed instance, shear-free bulk congruence tangent to $\partial_r$. The latter is part of the canonical null tetrad  parallelly transported  along  $\partial_r$ (thanks to the affine nature of $r$) introduced in \cite{CMPPS2}, which coincides with that of \cite{Stephani:624239}, Eq. (29.13a), as well as with the original Ref. \cite{NP68}. In complex celestial-sphere coordinates $\zeta$ and $\bar \zeta$, see App. \ref{carcot3}, the null tetrad reads:
\begin{equation}
\label{nultetrad}
\begin{cases}
\mathbf{k}=\partial_r \\
\mathbf{l}=\frac{1}{2}\left(\frac{8\pi G\varepsilon r+\ast \varpi c}{\rho^2}-r\theta-\hat{\mathscr{K}}\right)\partial_r+\upupsilon\\
\mathbf{m}=\frac{P}{r-\text{i}\ast\! \varpi}
\left(\hat\partial_{\bar \zeta}
+\left(\ast\hat{\mathscr{D}}_{\bar \zeta}\ast\!\varpi-r\varphi_{\bar \zeta}\right)\partial_r\right)
\end{cases}  
\end{equation}  
with the usual relations $\mathbf{k}\cdot\mathbf{l}=-1 $, $\mathbf{m}\cdot\bar{\mathbf{m}}=1$ and $\text{d}s^2_{\text{res. Ricci-flat}}=-2 \mathbf{k} \mathbf{l}+2 \mathbf{m}\bar{\mathbf{m}}$. Generically, $\mathbf{k}$ is a multiplicity-two principal null direction of the Weyl tensor, and using the tetrad at hand we find the following Weyl complex scalars:\footnote{Neither $\Uppsi_3$ nor $\Uppsi_4$ vanish in the instance of Petrov type D solutions, because $\mathbf{l}$ \emph{is not}  a principal null direction.  Another tetrad is reached  with a Lorentz transformation suitably adjusted for $\mathbf{l}' $ be a principal direction of multiplicity two whereas $\mathbf{k}'\propto \mathbf{k}$, and $\Uppsi_3'=\Uppsi_4'=0$. \label{typeD}}
\begin{equation}
\label{psis}
\begin{cases}
{\Uppsi}_0={\Uppsi}_1=0\\
{\Uppsi}_2=\frac{\text{i}\hat{\tau}}{2(r-\text{i}\ast\! \varpi)^3}\\
{\Uppsi}_3=\frac{\text{i}P\chi_\zeta}{(r-\text{i}\ast\! \varpi)^2}+ \text{O}\left(\nicefrac{1}{(r-\text{i}\ast\! \varpi)^3}\right)\\
{\Uppsi}_4=\frac{\text{i} X_\zeta^{\hphantom{\zeta}\bar \zeta}}{r-\text{i}\ast\! \varpi}+ \text{O}\left(\nicefrac{1}{(r-\text{i}\ast\! \varpi)^2}\right).
\end{cases}  
\end{equation}  
Unsurprisingly, all $\Uppsi$s are spelled using the Carrollian  descendants of the boundary Cotton tensor -- as well as their derivatives in the higher-order terms.

\subsection{Bulk versus boundary isometries}\label{bvbK}

The geometries under consideration possess at least one Killing vector field. A natural question to address concerns the boundary manifestation of a bulk isometry. At the same time such an analysis provides the recipe for designing bulk isometries from a purely boundary perspective.

We will circumscribe our investigation to vector fields, which have no component along $\partial_r$, and whose other components depend only on $t$ and $\mathbf{x}$.
We could be more general without much effort assuming e.g. an expansion in inverse powers of $r$ for the missing component and for the radial dependence of the others. However, this would unnecessarily sophisticate our presentation without shedding more light on our simple and robust conclusion: \emph{the bulk isometries at hand are mapped onto boundary Carrollian diffeomorphisms generated by strong Killing vectors} (a summary on Carrollian isometries is available in  App.~\ref{conscar}). 

It is convenient for the subsequent developments to adopt bulk Cartan frame and coframe aligned with the boundary \eqref{kert}, \eqref{dhat} and \eqref{kertdual}:
\begin{equation}
\label{bulkfrcofr}
\begin{array}{rcl}
& \text{e}_{\hat t}\equiv\upupsilon= \frac{1}{\Omega}\partial_t ,\quad  \text{e}_{\hat \imath}\equiv\hat\partial_i=\partial_i+\frac{b_i}{\Omega}\partial_t,\quad  \text{e}_{\hat r}\equiv \partial_r,
& \\
& \uptheta^{\hat t}\equiv-\upmu=\Omega \text{d}t -b_i \text{d}x^i ,\quad   \uptheta^{\hat \imath}\equiv\text{d}x^i,\quad  \uptheta^{\hat r}\equiv \text{d}r.&
\end{array}
\end{equation}
The components for the bulk metric \eqref{exp-ric-fl-res} read (in order to avoid cluttering, we keep the ``hat'' on the time indices only, where potential ambiguity exists): 
\begin{equation}
\label{resmetexp}
\begin{array}{rcl}
&g_{\hat t \hat t}= \frac{1 }{\rho^2}\left(8\pi G\varepsilon r+\ast \varpi c\right)-r\theta -\hat{\mathscr{K}}
,\quad
g_{\hat ti}= \ast\hat{\mathscr{D}}_{i}\ast\!\varpi  -r\varphi_i,\quad
g_{\hat tr}=-1, & \\
&g_{ri}= 0,\quad
g_{rr}=  0,\quad
g_{ij}= \rho^2 a_{ij}. &
\end{array}
\end{equation}
Assuming  a bulk vector of the form 
\begin{equation}   
\label{bulkil}
\upxi=\xi^t(t, \mathbf{x}) \partial_t +\xi^k(t, \mathbf{x}) \partial_k=\xi^{\hat t}(t, \mathbf{x}) \upupsilon +\xi^k(t, \mathbf{x}) \hat\partial_k,
\end{equation}  
where 
$\xi^{\hat t}=\Omega\xi^t -\xi^k b_k$, we can determine the Lie derivative of the metric:
\begin{equation}
\label{lieresmet}
\begin{array}{rcl}
&\mathscr{L}_\upxi g_{rr}=0,\quad
\mathscr{L}_\upxi g_{r\hat t}=\mu,\quad
\mathscr{L}_\upxi g_{ri}=\nu_i,&
\\
&\mathscr{L}_\upxi g_{ij}= 2 \rho^2 \left(\hat\nabla_{(i}\xi^ka_{j)k}+\xi^{\hat t}  \hat \gamma_{ij}\right) 
-2g_{\hat t(i}\nu_{j)}
+  a_{ij} \upxi\left(\ast\varpi^2\right),&
\\
&\mathscr{L}_\upxi g_{\hat ti}=- g_{\hat ti} \mu - g_{\hat t\hat t} \nu_i 
-r\left(\upxi\left(\varphi_i\right)+\varphi_j \hat\partial_i \xi^j\right)
+\upxi\left(\ast\hat{\mathscr{D}}_{i}\ast\!\varpi \right)+\left(\ast\hat{\mathscr{D}}_{j}\ast\!\varpi \right) \hat\partial_i \xi^j
+\rho^2 a_{ij}\frac{1}{\Omega}\partial_t\xi^j,&
\\
&\mathscr{L}_\upxi g_{\hat t\hat t}= -2g_{\hat t\hat t} \mu + 2 g_{\hat ti}\frac{1}{\Omega}\partial_t\xi^i-\upxi\left(
\frac{1 }{\rho^2}\left(8\pi G\varepsilon r+\ast \varpi c\right)-r\theta -\hat{\mathscr{K}}
\right)&
\end{array}
\end{equation}
with $\mu(t,\mathbf{x})$ and $\nu_i(t,\mathbf{x})$ given in \eqref{munui}. Observe that everything is expressed in terms of  
boundary Carrollian geometric objects (see App. \ref{carman}). 

Since the Killing components are $r$-independent, the above Lie derivative vanishes if and only if the coefficients of every power of $r$ do. The independent conditions we reach for this to occur are 
\begin{equation}
\label{cardifcon}
\partial_t\xi^i=0
\end{equation}
and \eqref{Carolisoa}, \eqref{CarolisoOm} and \eqref{super-K}, which therefore map the bulk Killing field \eqref{bulkil} onto a boundary Carrollian strong Killing vector (see App. \ref{conscar}). Some apparent extra conditions such as $ \upxi\left(\ast\varpi^2\right)=0$ or $\upxi\left(\varphi_i\right)+\varphi_j \hat\partial_i \xi^j=0$ are the vanishing of $\upxi$-Lie derivatives of some Carrollian tensors, which is guaranteed  by the strong Killing requirement on $\upxi$.


\subsection{Towers of charges and dual charges}\label{tc}

The Carrollian  dynamics emerging on the boundary as a consequence of bulk Einstein's equations, combined with the always available Carrollian conformal isometry group $\text{BMS}_4$, enables us to define a variety of charges. These are not necessarily conserved, but even in that instance, their evolution properties are canonical and provide an alternative, tamed picture of the dynamics. Furthermore, they should ultimately pertain to those charges recently discovered and discussed from a bulk perspective  \cite{Godazgar:2018vmm, Godazgar:2018qpq, Godazgar:2018dvh,Kol:2019nkc,Godazgar:2020gqd,Godazgar:2020kqd,Oliveri:2020xls,Kol:2020vet}, based as usual on the asymptotic symmetries -- also and unsurprisingly $\text{BMS}_4$, under appropriate fall-off conditions. Making the precise contact with those works would require a translation of our findings into the Newman--Penrose formalism \cite{NP68} beyond what we have already observed in Eqs. \eqref{psis}, namely ${\Uppsi}_2^0=\frac{\text{i}}{2}\hat \tau$,  ${\Uppsi}_3^0 =\text{i} P   \chi_\zeta$ and ${\Uppsi}_4^0 =\text{i}   X_{\zeta}^{\hphantom{\zeta}\bar\zeta}$. This would bring us far from our goal, and we will limit ourselves to pointing out that the ten Newman--Penrose conserved charges vanish here because the spacetimes are algebraically special. These charges would have been otherwise associated with the $s=1$ ``non-tilde'' class introduced below, involving non-zero $E_{ij}$ and $N_i$ in the non-algebraic instance.

Ricci-flat metrics, either in the general form \eqref{exp-ric-fl} or in its resummed version \eqref{exp-ric-fl-res}, exhibit two important features for the description of charges. Firstly, every order $\nicefrac{1}{r^{2s+1}}$ reveals Carrollian dynamics of the type  \eqref{carEbiscon}, \eqref{carFcon}, \eqref{carGcon} and \eqref{carHcon} with momenta  
$\Pi_{(s)}$, $\Pi_{(s)}^i$, $P_{(s)}^i$, $ \tilde\Pi_{(s)}^{ij}$ and $\Pi_{(s)}^{ij}$, and possibly with right-hand sides -- non-conservation. Every such set of momenta together with the Carrollian conformal Killings \eqref{Xclock} lead to currents $\kappa_{(s)}$, $K^i_{(s)}$, $\tilde \kappa_{(s)}$, $\tilde K^i_{(s)}$ and charges
$Q_{(s)\, T,Y}$ and $\tilde Q_{(s)\, T,Y}$, following  \eqref{carconch} and  \eqref{carkappaK}. Their conservation or evolution encoded in \eqref{carconchdt} depends on 
 $\mathcal{K}_{(s)}$, $\tilde{\mathcal{K}}_{(s)}$ in \eqref{carKN}. The set associated with $s=0$ corresponds to the fluid momenta \eqref{mom-fluid} and its charges are \emph{leading}; the sets with $s\geq1$ reveal the \emph{subleading charges}.  Moreover, all these charges should be referred to as \emph{electric} because their conservation, if valid, occurs on-shell.
  
 Secondly,  the Carrollian Cotton tensors obey conservation equations \eqref{carEbiscon}, \eqref{carFcon}, \eqref{carGcon} and \eqref{carHcon} with momenta \eqref{mom-Cot}, leading to two towers of Cotton charges $Q_{\text{Cot}\, T,Y}$ and $\tilde Q_{\text{Cot}\, T,Y}$, as discussed in App. \ref{carcot3}. These charges are \emph{magnetic} as the conservation of the Cotton is an identity valid off-shell.\footnote{We borrow here the phrasing \emph{electric} and \emph{magnetic} from Refs.   \cite{Ashtekar82, Bossard:2008sw}.}  Furthermore, the Carrollian Cotton tensors are not exclusive to $\nicefrac{1}{r^{2}}$: each order $\nicefrac{1}{r^{2s+2}}$ brings its share of off-shell Carrollian dynamics  with momenta $\Pi_{\text{Cot}\,(s)}$, $\Pi_{\text{Cot}\,(s)}^i$, $P_{\text{Cot}\,(s)}^i$, $ \tilde\Pi_{\text{Cot}\,(s)}^{ij}$, $\Pi_{\text{Cot}\,(s)}^{ij}$, currents $\kappa_{\text{Cot}\,(s)}$, $K^i_{\text{Cot}\,(s)}$, $\tilde \kappa_{\text{Cot}\,(s)}$, $\tilde K^i_{\text{Cot}\,(s)}$, and finally magnetic charges $Q_{\text{Cot}\,(s)\, T,Y}$ and $\tilde Q_{\text{Cot}\,(s)\, T,Y}$.

Incidentally, it should be noticed that due to the relationships amongst the fluid and the Cotton (Eq. \eqref{visc-cargen-resum-expli} in general plus Eq. \eqref{pi-fluid-resum} in the resummable family), the electric and the magnetic towers have a non-empty intersection: $\tilde Q_{(s)\, T,Y}$ and $\tilde Q_{\text{Cot}\,(s)\, T,Y}$ 
generally 
coincide.

Let us for concreteness overview the situation in the resummable instance, Eq. \eqref{exp-ric-fl-res}. Expanding the resummed factor $\nicefrac{1}{\rho^2}$, we find the following 
results. 
\begin{description}
\item[Electric towers] These have $s$th momenta $\Pi_{(s)}$, $\Pi_{(s)}^i$, $P_{(s)}^i$, $ \tilde\Pi_{(s)}^{ij}$ and $\Pi_{(s)}^{ij}$ equal to \eqref{mom-fluid-res} multiplied by $\ast \varpi^{2s}$. The same factor will multiply the leading Carrollian current ($s=0$ i.e. $\kappa$, $K^i$, $\tilde \kappa$, $\tilde K^i$) and give the $s$th, $\kappa_{(s)}$, $K^i_{(s)}$, $\tilde \kappa_{(s)}$ and $\tilde K^i_{(s)}$, following \eqref{carkappaK}. Using the Carroll--Bianchi identities  \eqref{Carroll-Bianchi1}, \eqref{Carroll-Bianchi2} and \eqref{Carroll-Bianchi3}, we find the divergences \eqref{carKN}, which contribute the time evolution of the charges computed as in  \eqref{carconch}, using  \eqref{carconchdt}:
\begin{eqnarray}
\tilde{\mathcal{K}}_{(s)}&=&-s\ast\!\varpi^{2s-2}\left(\ast\varpi \hat{\mathscr{A}}\tilde \kappa+\frac{1}{3} \tilde K^i \ast\!\psi_i\right),
\\
\mathcal{K}_{(s)}&=&-\frac{\ast\varpi^{2s}}{8\pi G} \ast \! \chi^{i}\left(\hat{\mathscr{D}}_i\xi^{\hat t }-
2\xi^j \varpi_{ji}
\right)-s\ast\!\varpi^{2s-2}\left(\ast\varpi \hat{\mathscr{A}}\kappa+\frac{1}{3}  K^i \ast\!\psi_i\right)
\label{dives}
\end{eqnarray}
with
\begin{equation}
\label{carkappaK-res}
 \begin{cases}
\kappa=  \frac{1}{8\pi G }\xi^{i}\ast\!\psi_i-\xi^{\hat t}  \varepsilon
\\
\tilde\kappa=\frac{1}{8\pi G} \xi^{i} \ast \! \chi_{i}
\\
K^i=\frac{\varepsilon}{2}\xi^{i}-\frac{1}{8\pi G}\left(\xi^{j}\ast  \!  \Psi^{i}_{\hphantom{i}j}+\xi^{\hat t} \ast \! \chi^{i}\right)
\\
\tilde K^i =-\frac{1}{8\pi G} \xi^{j}\ast \! X^{i}_{\hphantom{i}j},
\end{cases}
\end{equation}
and the Killing components $\xi^{\hat t}$ and $\xi^i$ read off in \eqref{Xclock} following \eqref{carkil}.

Regarding the charges and their evolution, only  $\tilde Q_{(0)\, T,Y}=\int_{\mathscr{S}}\text{d}^{2}x \sqrt{a}\left(\tilde \kappa+b_j\tilde K^j\right)\equiv  \tilde Q_{T,Y} $ are \emph{always} conserved. These charges are purely geometric because they are integrals over $\mathscr{S}$\footnote{We use the property $V^i\ast\!W_i=-\ast\!V^iW_i$ -- see \eqref{hodgeast}.}
\begin{equation}
 \tilde Q_{T,Y} =
-\frac{1}{8\pi G} \int_{\mathscr{S}}\text{d}^{2}x \sqrt{a} \ast \! \xi^{i}\left(
\chi_{i}-b_jX^{j}_{\hphantom{j}i}
\right)
,
  \label{tildeintcarconch}
 \end{equation}
which do not involve the energy density $\varepsilon$, as opposed to $ Q_{(0)\, T,Y}=\int_{\mathscr{S}}\text{d}^{2}x \sqrt{a}\left( \kappa+b_j K^j\right) \equiv Q_{T,Y} $ spelled as
\begin{equation}
 Q_{T,Y} =
-\frac{1}{8\pi G}\int_{\mathscr{S}}\text{d}^{2}x \sqrt{a}\xi^{\hat t} \left(8\pi G\varepsilon+b_i \ast\!\chi^i\right)+\frac{1}{8\pi G}\int_{\mathscr{S}}\text{d}^{2}x \sqrt{a}
\xi^i\left( \ast\psi_i+4 \pi G\varepsilon b_i-b_j \ast\!\Psi^{j}_{\hphantom{j}i}\right)
.
  \label{intcarconch}
 \end{equation}
The latter are conserved for strong Carrollian Killings. Other charges might also be conserved for specific Carrollian conformal Killings, or depending on the configuration.

\item[Magnetic towers] The $s$th magnetic momenta $\Pi_{\text{Cot}\,(s)}$, $\Pi_{\text{Cot}\,(s)}^i$, $P_{\text{Cot}\,(s)}^i$, $ \tilde\Pi_{\text{Cot}\,(s)}^{ij}$ and $\Pi_{\text{Cot}\,(s)}^{ij}$ are  \eqref{mom-Cot} multiplied by $\ast \varpi^{2s}$. As for the electric case, this latter factor will appear in all magnetic currents $\kappa_{\text{Cot}\,(s)}$, $K^i_{\text{Cot}\,(s)}$, $\tilde \kappa_{\text{Cot}\,(s)}$ and $\tilde K^i_{\text{Cot}\,(s)}$ built out of the leading $s=0$:
\begin{equation}
\label{carkappaK-cot-res} 
 \begin{cases}
\kappa_{\text{Cot}}=  \xi^{i}\psi_i-\xi^{\hat t} c
\\
\tilde\kappa_{\text{Cot}}=\xi^{i} \chi_{i}
\\
K_{\text{Cot}}^i=\frac{c}{2}\xi^{i}-\xi^{j} \Psi^{i}_{\hphantom{i}j}-\xi^{\hat t} \chi^{i}
\\
\tilde K_{\text{Cot}}^i =- \xi^{j} X^{i}_{\hphantom{i}j},
\end{cases}
\end{equation}
Their divergences \eqref{carKN} read: 
\begin{eqnarray}
\tilde{\mathcal{K}}_{\text{Cot}\,(s)}&=&-s\ast\!\varpi^{2s-2}\left(\ast\varpi \hat{\mathscr{A}}\tilde \kappa_{\text{Cot}}+\frac{1}{3} \tilde K_{\text{Cot}}^i \ast\!\psi_i\right),
\\
\mathcal{K}_{\text{Cot}\,(s)}&=&-\ast\!\varpi^{2s}\chi^{i}\left(\hat{\mathscr{D}}_i\xi^{\hat t }-
2\xi^j \varpi_{ji}
\right)-s\ast\!\varpi^{2s-2}\left(\ast\varpi \hat{\mathscr{A}}\kappa_{\text{Cot}}+\frac{1}{3}  K_{\text{Cot}}^i \ast\!\psi_i\right).
\label{divms}
\end{eqnarray}
These determine the evolution \eqref{carconchdt} of the charges \eqref{carconch}, from which we learn that $\tilde Q_{\text{Cot}\,(0)\, T,Y}=\int_{\mathscr{S}}\text{d}^{2}x \sqrt{a}\left(\tilde \kappa_{\text{Cot}}+b_j\tilde K_{\text{Cot}}^j \right)\equiv \tilde Q_{\text{Cot}\,T,Y}$ are always conserved:
\begin{equation}
\tilde Q_{\text{Cot}\,T,Y}=
\int_{\mathscr{S}}\text{d}^{2}x \sqrt{a}\xi^{i}\left(
\chi_{i}-b_jX^{j}_{\hphantom{j}i}
\right).
  \label{tildeintcarconchcot}
 \end{equation}
For strong Carrollian Killing fields,  
$Q_{\text{Cot}\,(0)\, T,Y}=\int_{\mathscr{S}}\text{d}^{2}x \sqrt{a}\left(\kappa_{\text{Cot}}+b_j K_{\text{Cot}}^j \right)\equiv Q_{\text{Cot}\,T,Y}$ given by 
\begin{equation}
Q_{\text{Cot}\,T,Y}=
-\int_{\mathscr{S}}\text{d}^{2}x \sqrt{a}\xi^{\hat t} \left(c+b_i \chi^i\right)+\int_{\mathscr{S}}\text{d}^{2}x \sqrt{a}
\xi^i\left(\psi_i+\frac{c}{2}b_i-b_j\Psi^{j}_{\hphantom{j}i}\right)
  \label{intcarconchcot}
 \end{equation}
are also conserved off-shell, as other magnetic charges are in specific situations. 

\end{description}
Several comments are in order here concerning the above sets of charges obtained for the resummable metrics \eqref{exp-ric-fl-res}. The tower of electric geometric charges $\tilde Q_{(s)\, T,Y}$, constructed upon multiplying the integrand of  \eqref{tildeintcarconch} by $\ast \varpi^{2s}$,
coincides with its magnetic counterpart $\tilde Q_{\text{Cot}\,(s)\, T,Y}$ obtained  likewise using \eqref{tildeintcarconchcot}.
 In $d=2$, if $\xi^i$ are the spatial components of a conformal Killing field, so are $\ast\xi^i$.\footnote{The proof of this statement is straightforward in complex coordinates, see footnote \ref{astkil}.} Hence the set of all $\xi^i$s is identical to that of $\ast\xi^i$s. The associated charges could be called ``self-dual,'' and in total three distinct towers emerge:  the self-dual $\left\{\tilde Q_{(s)\, T,Y}\right\}\equiv\left\{ \tilde Q_{\text{Cot}\,(s)\, T,Y}\right\}$,  the electric $\left\{Q_{(s)\, T,Y}\right\}$ and the magnetic $\left\{Q_{\text{Cot}\,(s)\, T,Y}\right\}$ -- the last two are reached by inserting $\ast \varpi^{2s}$ into the integrals  \eqref{intcarconch} and  \eqref{intcarconchcot}.
 The $\ast \varpi^{2s}$ insertion pattern  grants the subleading towers with the status of \emph{multipolar moments} (see the original works  \cite{  GerochMI, GerochMII, Hansen, Fodor
}  as well as \cite{Compere:2017wrj} for a modern perspective). Making this statement precise would force us to deviate substantially from the analysis of the hidden M\"obius group. This could fit more naturally in a comprehensive comparison of the present approach to subleading charges with the  rich existing literature quoted earlier. Nonetheless, the pertinence of the proposition will be illustrated in the example of Kerr solution, at the very end of the forthcoming section \ref{static}. 

Among the above towers of $\text{BMS}_4$ charges, always present but not always conserved, one finds those corresponding to the  
bulk isometries, whenever present. Indeed, as discussed in Sec. \ref{bvbK}, bulk Killings of the form \eqref{bulkil} are associated with boundary strong Carrollian Killing vector fields. Combined as previously with the leading and subleading, electric and magnetic momenta, they generate two electric and two magnetic towers of charges: $\left\{ Q_{(s)}, \tilde Q_{(s)}, Q_{\text{Cot}\,(s)},  \tilde Q_{\text{Cot}\,(s)}\right\}$. The four leading charges $\left\{ Q_{(0)}, \tilde Q_{(0)}, Q_{\text{Cot}\,(0)},  \tilde Q_{\text{Cot}\,(0)}\right\}$ are always conserved, but part of them may be trivial or not independent. 
The subleading are neither necessarily conserved, nor always independent, and have the status of electric and magnetic multipole moments.

\subsection{Time-independent solutions}\label{static}

\subsubsection*{Reconstruction from the boundary}

In view of the forthcoming Ehlers--Geroch reduction, we will now assume the existence of a time-like Killing vector field $\upxi$ in Ricci-flat solutions of the resummable type \eqref{exp-ric-fl-res}. Such a vector could be generally of the form \eqref{bulkil}. 
In stationary spacetimes, the field $\upxi$ remains time-like in the asymptotic region. Then, it is  possible to choose the field $\upupsilon$ \eqref{kert} of the modified Newman--Unti gauge such that $\upupsilon\equiv\upxi$. Setting further $\Omega=1$ brings  the Killing to the simple form  $\partial_t$ (see e.g. \cite{NP68} for a detailed description of the procedure). On the conformal boundary, the time-like Killing congruence thus coincides with the fibre of the Carrollian bundle. This feature is absent for spacetimes where a time-like Killing  field exists but becomes space-like in the asymptotic region. Examples of this sort are captured by the Pleba\'nski--Demia\'nski family (like the C-metric)  \cite{Plebanski:1976gy} (see also \cite{Stephani:624239, Podolsky}), which is algebraically special of Petrov type D.\footnote{Their Weyl components are given in Eq. \eqref{psis} -- see also footnote \ref{typeD}.}
These  include the black-hole acceleration parameter, which is responsible for the appearance of another Killing horizon, creating a new asymptotic region where the Killing vector fails to be time-like. 
 Although interesting on its own right -- of limited physical use, however -- the inclusion of this parameter would render the presentation too convoluted,  in particular because the action of the Ehlers group in this instance does not respect the algebraic feature of the spacetime. For the sake of clarity we will restrict our investigation to Killings of the form $\partial_t$, aligned with the fiber, i.e. to truly stationary spacetimes, which remain algebraically special under Ehlers transformations.

With the present choice, none of the Carrollian building blocks of $\text{d}s^2_{\text{res. Ricci-flat}}$ depends on $t$. As a consequence (see Apps. \ref{carman} and \ref{carcot3}) $\theta=0$ and $\varphi_i=\partial_i \ln \Omega$. The latter can be set to zero with a time-independent Weyl rescaling, which therefore amounts to setting $\Omega=1$. This is an innocuous gauge fixing that will be assumed here because it allows to severely simplify the dynamics. Backed with time independence, Carrollian Weyl-covariant derivatives become ordinary 
Levi--Civita derivatives, and the only non-vanishing tensors are the following, in complex coordinates with $P=P(\zeta,\bar \zeta)$ -- see App. \ref{carcot3}: 
\begin{eqnarray}                           
\label{astvpitinv}
&\ast \varpi=\dfrac{\text{i}P^2}{2}\left(
\partial_{\zeta}b_{\bar\zeta}-\partial_{\bar\zeta} b_{\zeta}
\right),&\\
\label{Ktinv}
&\hat{\mathscr{K}}=\hat{K}=K=\Delta \ln P,&
\end{eqnarray}
\begin{eqnarray}                           
\label{c-Carrolholtinv}
&c=\left(\Delta +2K\right)\ast\! \varpi,&
\\
\label{carcotch}
& \chi_{\zeta}=\frac{\text{i}}{2}\partial_{\zeta}K, \quad 
\chi_{\bar\zeta}=-\frac{\text{i}}{2}\partial_{\bar\zeta}K,&\\
\label{carcotps}
&\psi_{\zeta}=3\text{i}\partial_{\zeta}\ast\! \varpi^2,\quad 
 \psi_{\bar\zeta}=-3\text{i}\partial_{\bar\zeta}\ast\! \varpi^2,&\\
\label{carcotPs}
& \Psi_{\zeta\zeta}=
\dfrac{1}{P^2}\partial_{\zeta}\left(P^2\partial_{\zeta}
\ast\! \varpi \right), \quad
\Psi_{{\bar\zeta}{\bar\zeta}}=
\dfrac{1}{P^2}\partial_{\bar\zeta}\left(P^2\partial_{\bar\zeta}
\ast\! \varpi \right),
\end{eqnarray}
where $\Delta f= 2P^2 \partial_{\bar\zeta} \partial_{\zeta}f $. To these one should add the energy density (i.e. the Bondi mass aspect) $\varepsilon$, as well as another scalar
\begin{equation}
\label{astdvpitinv}
\varpi=\dfrac{P^2}{2}\left(
\partial_{\zeta}b_{\bar\zeta}+\partial_{\bar\zeta} b_{\zeta}
\right),
\end{equation}
which is  $\frac{1}{2} \nabla_i b^i$ and should not be confused with the two-form $\upvarpi=\frac{1}{2}\varpi_{ij}\text{d}x^i\wedge \text{d}x^j$, i.e. the Hodge-dual of the scalar $\ast \varpi = - \frac{1}{2} \nabla_i \ast\! b^i$ displayed explicitly in \eqref{astvpitinv}. These two real twist scalars are adroitly combined into the \emph{complex Carrollian twist}
\begin{equation}
\label{astdvpitinvc}
\hat \varpi= \ast \varpi+\text{i} \varpi.
 \end{equation}

The equations of motion \eqref{carEbiscon-res}, \eqref{carGconres} (or in the form \eqref{divtau},  \eqref{lapltau} with $\hat \tau$ defined in \eqref{taudef}) are recast as 
\begin{eqnarray}
\label{Kharm}
\Delta K &=&0,\\
\label{charm}
\partial_\zeta \hat \tau&=&0.
 \end{eqnarray}
The first shows  that the curvature is required to be a harmonic function i.e.
\begin{equation}
\label{k}
 K(\zeta, \bar\zeta ) =\frac{1}{2}\left(\hat k(\zeta)+\hat{\bar k}(\bar\zeta )\right),
 \end{equation}
and although $\hat k(\zeta)$ is an arbitrary holomorphic function, the freedom is rather limited as $K$ must also be the Laplacian of $\ln P$. Besides the constant-curvature cases, one solution has been exhibited thus far \cite{Stephani:624239} (up to holomorphic coordinate transformations): $K=-3(\zeta+\bar \zeta)$ realized with $P=(\zeta+\bar \zeta)^{\nicefrac{3}{2}} $. We will not specify any particular choice for the moment. For future use, we define the imaginary part of $\hat k(\zeta)$ as another harmonic function
\begin{equation}
\label{kst}
 K^\ast(\zeta, \bar\zeta ) =\frac{1}{2\text{i}}\left(\hat k(\zeta)-\hat{\bar k}(\bar\zeta )\right).
 \end{equation}
From Eqs. \eqref{charm} and \eqref{taudef}, we infer that $-c $ is the real part of an arbitrary holomorphic function $\hat \tau(\zeta)$, whereas the imaginary part of the latter is $8\pi G \varepsilon$; both are harmonic functions. Given $c$ and $K$, we can proceed with Eq. \eqref{c-Carrolholtinv} and find $\ast \varpi$, from which it is always possible to determine $b_\zeta $ and $b_{\bar \zeta}$.

Although the focus of the present work is not to solve Einstein's equations, we will elaborate for illustrative purposes on the steps we've just described, without delving into fine questions like completeness or gauge redundancy of the solutions. Note in passing how remarkably the Carrollian boundary formalism is adapted to the framework of Ricci-flat spacetimes, allowing to convey often complicated expressions in a very elegant manner, and sorting naturally otherwise scattered classes of solutions (the ones we present can be found in various chapters of Refs. \cite{Stephani:624239, Podolsky}). Several distinct instances appear, which require a separate treatment.
\begin{description}
\item[Non-constant $\pmb K$] 

This is the generic situation, although in practice the most obscure regarding the interpretation of the bulk geometries. As already mentioned, very few $P$s are expected to possess a non-constant harmonic curvature $K$, but assuming one has one, accompanied by its holomorphic function $\hat k(\zeta) $, and making a choice for the arbitrary holomorphic function $\hat \tau(\zeta) $, Eq.  \eqref{c-Carrolholtinv} can be solved for $\ast \varpi$, which is expressed using \eqref{astvpitinv} with Ehresmann connection
\begin{equation}
\label{bz-nonck}
b_\zeta(\zeta, \bar\zeta)=\frac{ \text{i}\hat{\bar \tau}( \bar\zeta)}{P^2(\zeta, \bar\zeta)\partial_{\bar\zeta}\hat{\bar k}( \bar\zeta)}
.
 \end{equation}

\item[Constant $\pmb K$] This implies that  $\hat k(\zeta) $ is also constant and the above solution is invalid. The situation at hand is the most common, however, as it captures three standard instances: spherical, flat or hyperbolic foliations. We can parameterize the function $P$ as follows:
 \begin{equation}
\label{P-conK}
P(\zeta, \bar\zeta)= A \zeta \bar\zeta +B \zeta +\bar B \bar\zeta + D
 \end{equation}
with $A$, $D$ arbitrary real constants and $B$ an arbitrary complex constant, leading to
 \begin{equation}
\label{conK}
K= 2(A D -B \bar B ).
 \end{equation}
 Several cases emerge, which must be treated separately.
\begin{description}
\item[$\pmb{K\neq 0}$] Here $c(\zeta,\bar\zeta ) =-\frac{\hat \tau(\zeta) + \hat{\bar \tau}(\bar\zeta)}{2}$ is an arbitrary (possibly constant) harmonic function, and Eq.  \eqref{c-Carrolholtinv} is solved with
\begin{equation}
\label{astvp-ck}
\ast\varpi(\zeta, \bar\zeta)=\frac{c(\zeta, \bar\zeta)}{2K}
+
\text{i}\left(\bar f(\bar\zeta)\partial_{\bar\zeta}\ln P(\zeta, \bar\zeta)
-f(\zeta)\partial_{\zeta}\ln P(\zeta, \bar\zeta)
+\frac12\left(
\partial_{\zeta}f(\zeta)
-\partial_{\bar\zeta}\bar f(\bar\zeta)\right)
\right)
 \end{equation}
 with $f(\zeta)$ an arbitrary holomorphic function. It is reached with the following Ehresmann connection ($ \hat{\tau}_0$ is a real constant):
\begin{equation}
\label{bz-ck}
b_\zeta(\zeta, \bar\zeta)=\displaystyle{-\frac{ \bar \zeta\left( \hat{\tau}_0+ \text{i} \hat{\tau}(\zeta)\right)}{2 K (B\zeta +D) P(\zeta, \bar\zeta) }
+ \frac{ \bar f(\bar\zeta)}{P^2(\zeta, \bar\zeta) }}.
\end{equation}


\item[$\pmb{K=0}$]  This instance is obtained with $A=B=0$ so that $P=D$. Now, given an arbitrary harmonic function $c(\zeta, \bar\zeta)=-\frac{\hat \tau(\zeta) + \hat{\bar \tau}(\bar\zeta)}{2}$ and an arbitrary holomorphic function $Z(\zeta)$, we find
\begin{equation}
\label{astvp-K0}
\ast\varpi(\zeta, \bar\zeta)=\frac{\text{i}}{2}\left(
Z(\zeta)
-\bar Z(\bar\zeta)\right)
-\frac{1}{4P^2}\left(
\bar \zeta\int^\zeta \text{d}z\,  \hat\tau(z)+
 \zeta\int^{\bar\zeta} \text{d}\bar z\,  \hat{\bar\tau}(\bar z)
\right)
,
 \end{equation}
and ($ \hat{\tau}_0$ is a real integration constant)
\begin{equation}
\label{bz-K0}
b_\zeta(\zeta, \bar\zeta)=\displaystyle{\frac{1}{P^2}\int^{\bar\zeta} \text{d}\bar z\,  \bar Z(\bar z)-\frac{\bar \zeta^2}{4P^4}\int^\zeta \text{d}z\left(\hat{\tau}_0+ \text{i} \hat\tau(z)\right)}.
 \end{equation}

\end{description}
The last two cases have in common the instance where $c=K=0$, realized with vanishing $\hat \tau$ and constant $P$.

\end{description}

As already noticed, all solutions described in a unified fashion here can be found in the earlier quoted literature under distinct labels.\footnote{It should be stressed that part of the present solution space originates in gauge freedom. In particular, $c(\zeta,\bar\zeta)$ being Weyl-covariant of weight $3$ (see App. \ref{carcot3}), it can always be reabsorbed by a boundary Weyl transformation, which is in turn neutralized by a bulk $r$-rescaling. Such a boundary transformation will bring $\Omega$ back with non-vanishing  $\varphi_i$, which we have set to zero, and this is the reason we cannot here restrict to constant $c$ and $\hat\tau$.} Discussing them would take us outside of our objectives. We will only emphasize a notorious subclass, which is the Kerr--Taub--NUT family. For the latter, the curvature $K$ is constant \eqref{conK} and realized e.g. with $B=0$. Two distinct instances emerge: vanishing and non-vanishing $K$, respectively obtained with vanishing and non-vanishing $A$.
\begin{itemize}
\item For non-vanishing $K$,  the holomorphic function $\hat \tau$ is 
\begin{equation}
\label{tau-TN}
\hat \tau =2\text{i}(M+\text{i}K n),
\end{equation}
where $M$ is the mass and $n$ the nut charge, both constants. The holomorphic function $f(\zeta)$ reads 
\begin{equation}
\label{f-TN}
f(\zeta) =\text{i} a \zeta
\end{equation}
with $a$ the Kerr angular velocity. Using Eqs. \eqref{astvp-ck} and \eqref{bz-ck} with $ \hat{\tau}_0=2M$ we find:
\begin{equation}
\label{bz-ckKTN}
b_\zeta(\zeta, \bar\zeta)=-\text{i} \bar\zeta\left(
\frac{a}{P^2}-\frac{ n}{DP}
\right)
\end{equation}
and
\begin{equation}
\label{astvp-ckKTN}
\ast\varpi(\zeta, \bar\zeta)=n+a-\frac{2Da}{P},
\end{equation}
where $P=A\zeta\bar\zeta+D$
and $K=2AD$.
\item For  $K=0$ (i.e. $P=D$ constant), we use Eqs. \eqref{astvp-K0} and   \eqref{bz-K0} with $ \hat{\tau}_0=2M$,\footnote{Both for vanishing and non-vanishing $K$, $ \hat{\tau}_0$ has been tuned to ensure that $M$ does not appear in $b_\zeta$, displayed in \eqref{bz-ckKTN} and \eqref{bz-K0KTN}. There is no principle behind this choice, it is simply in line with standard conventions  for the Kerr--Taub--NUT family. As a consequence, $\varpi$ defined in \eqref{astdvpitinv} vanishes.
\label{tau0}}
\begin{equation}
\label{tau-TN-K0}
\hat \tau =2\text{i}M
\end{equation}
and 
\begin{equation}
\label{Z-TN}
Z =\text{i}a.
\end{equation}
This leads to 
\begin{equation}
\label{bz-K0KTN}
b_\zeta(\zeta, \bar\zeta)=-\text{i}\frac{\bar\zeta a}{P^2}
\end{equation}
and
\begin{equation}
\label{astvp-K0KTN}
\ast\varpi=-a.
\end{equation}
Observe the absence of nut charge in the present case.\footnote{Despite the absence of magnetic charges, the solution at hand belongs formally to the Taub--NUT family (see Ref. \cite{Podolsky}, \S 12.3.2).

}

\end{itemize}

\subsubsection*{A remark on the rigidity theorem}

The \emph{rigidity theorem} asserts that under appropriate hypotheses, the isometry group of stationary asymptotically flat spacetimes contains $\mathbb{R}\times U(1)$. This theorem is best presented in Refs. \cite{Chrusciel:1996bm,Chrusciel:1996bj}, where the necessary assumptions are stated more accurately than in the original discussions (see e.g. \cite{HE}). Our framework does embrace stationary spacetimes. However, we have been agnostic regarding analyticity or regularity properties, which turn out to be fundamental for the applicability of the theorem at hand. Hence, we have no reason to foresee any additional $U(1)$ symmetry in all reconstructed solutions of the present chapter. 

Aside from mathematical rigor, we can recast the conceivable disruption of the rigidity theorem from the boundary perspective, which has been our viewpoint. We have shown in Sec. \ref{bvbK} that a bulk Killing field is mapped onto a Carrollian strong Killing on the boundary. The generator of the desired  $U(1)$ is of the form \eqref{bulkil} with no time leg\footnote{We could keep non-vanishing $\xi^t$ and perform a more thorough analysis. This would not alter the conclusions, which are meant here to illustrate possible boundary faults in the rigidity theorem.} i.e. $\xi^t = 0$, and no time dependence in $\xi^i$ as imposed by \eqref{cardifcon}: 
\begin{equation}
\upxi=-\left(\xi^\zeta b_\zeta+\xi^{\bar \zeta} b_{\bar \zeta}\right)\partial_t +\xi^\zeta \hat\partial_\zeta+\xi^{\bar \zeta}  \hat\partial_{\bar\zeta}.
\end{equation}
A strong Carrollian Killing field must obey  Eqs.  \eqref{Carolisoa}, \eqref{CarolisoOm} and \eqref{super-K}. Here \eqref{CarolisoOm} is identically satisfied, whereas \eqref{Carolisoa} leads to 
\begin{equation}
\label{CarolisoOmtind} 
\partial_\zeta \xi^{\bar \zeta}  =\partial_{\bar \zeta} \xi^\zeta=0,\quad \partial_\zeta \frac{\xi^\zeta}{P^2}  +\partial_{\bar \zeta} \frac{\xi^{\bar \zeta}}{P^2} =0.
\end{equation}
Finally \eqref{super-K} reads:
\begin{equation}
 \label{super-K-ntd} 
 P^2\partial_\zeta\left(\xi^\zeta b_\zeta+\xi^{\bar \zeta} b_{\bar \zeta}\right)+2\text{i}\, \xi^{\bar\zeta} \ast\! \varpi =0
\end{equation}
plus its complex conjugate.

For arbitrary $P(\zeta,\bar \zeta)$, Eqs. \eqref{CarolisoOmtind} have no solution, hence no extra Killing field is available. As mentioned earlier in the present section, the $P$s with harmonic curvature (required in \eqref{Kharm}) are very restricted and probably lack the necessary analyticity properties, explaining why the rigidity theorem is not applicable. This indeed happens in the quoted example with $P=(\zeta+\bar \zeta)^{\nicefrac{3}{2}} $.

Alternatively, considering $P=A\zeta\bar\zeta+D$ with constant curvature $K=2AD$, we find three more solutions to the equations \eqref{CarolisoOmtind}:
\begin{eqnarray}
\label{XK}
\upxi_1&=&\text{i}\left(\zeta \partial_\zeta-
\bar\zeta \partial_{\bar\zeta}
\right),\\
\label{YK}
\upxi_2&=&\frac{\text{i}}{2\sqrt{\vert AD\vert}}\left(\left(D-A\zeta^2\right)
\partial_\zeta-
\left(D-A\bar\zeta^2\right)\partial_{\bar\zeta}
\right),\\
\label{ZK}
\upxi_3&=&\frac{1}{2\sqrt{\vert AD\vert}}\left(\left(D+A\zeta^2\right)
\partial_\zeta+
\left(D+A\bar\zeta^2\right)\partial_{\bar\zeta}\right),
\end{eqnarray}
closing in $\mathfrak{so}(3)$,  $\mathfrak{e}_2$ and $\mathfrak{so}(2,1)$ algebras\footnote{The Lie brackets of the $\upxi$s are 
$\left[\upxi_1, \upxi_2\right]=\upxi_3$, 
$\left[\upxi_3, \upxi_1\right]=\upxi_2$ and $\left[\upxi_2, \upxi_3\right]=\frac{K}{\vert K\vert}\upxi_1$. For vanishing $K$, $\upxi_2= 
\frac{\text{i}}{2}\left(
\partial_\zeta-\partial_{\bar\zeta}
\right)
$ and $\upxi_3= 
\frac{1}{2}\left(
\partial_\zeta+\partial_{\bar\zeta}
\right)
$ 
are the translation commuting generators of $\mathfrak{e}_2$.
} for positive, zero or negative $K$.
Using \eqref{bz-ckKTN} and \eqref{astvp-ckKTN} one shows that for generic angular velocity $a$ and nut charge $n$,
only $\upxi_1$ obeys the strong condition  \eqref{super-K-ntd}. This is then promoted to a bulk field generating the rotational $U(1)$ isometry of the Kerr--Taub--NUT family. For vanishing $a$ and $n$, all three Carrollian Killing fields are strong and the bulk Ricci flat solution is fully isotropic -- Schwarzschild or A-class metric, see \cite{Stephani:624239, Podolsky}.

\subsubsection*{Charge analysis}

We would like to close the present section with a brief account on the charges of the Ricci-flat solutions under investigation.
Gravitational charges disclose the identity of a background and, as we have proposed in Sec. \ref{tc}, boundary Carrollian geometry supplies alternative techniques for their determination and the study of their conservation. These techniques are still in an incipient stage though, because the contact with the standard methods still needs to be elaborated. Furthermore, non-radiating configurations, in particular stationary and algebraically special, offer a limited playground in this programme. We would like nevertheless to summarize the situation, in view of the follow-up discussion on M\"obius hidden-group action, Sec. \ref{behch}. 

The simplest non-vanishing charge is the electric curvature defined in \eqref{elmgcarconch}:\footnote{Remember that here $\xi_{ij}=0$, and the geometry is $t$-independent with vanishing  $\theta$, $\varphi_i$, $ \hat{\mathscr{A}}$, $ \hat{\mathscr{R}}_i$ as well as $X_{ij}$.}
\begin{equation}
\label{Qelstat}
Q_{\text{ec}}=\int_{\mathscr{S}}\frac{\text{d}\zeta \wedge \text{d}\bar \zeta}{\text{i}P^2}K.
\end{equation}
Divided by the volume of $\mathscr{S}$, this is simply the average Gauss curvature. Note in passing that the charges defined here are extensive, hence the integrals may reveal convergence issues, in particular when $\mathscr{S}$ is non-compact. Normalizing with $\text{Vol}=\int_{\mathscr{S}}\frac{\text{d}\zeta \wedge \text{d}\bar \zeta}{\text{i}P^2}$ is the simplest way to fix this divergence.\footnote{The integrals can be performed by setting $\zeta=Z \text{e}^{i\varPhi}$, where $0\leq\varPhi< 2\pi$ and 
$Z=\sqrt{2}\tan \frac{\varTheta}{2}$, ${0<\varTheta<\pi}$ for $\mathbb{S}^2$;
$Z=\frac{R}{\sqrt{2}}$, $0<R<+\infty$ for $\mathbb{E}_2$;  $Z=\sqrt{2}\tanh \frac{\Psi}{2}$,  
$0<\Psi<+\infty$ for $\mathbb{H}_2$.}
Alternatively, $\mathscr{S}$ could be compactified -- quotiented by a discrete isometry group. We will leave this discussion aside, as it would be better addressed within attempts to make sense of Ricci-flat black holes with non-compact horizons (see e.g. Ch. 9 of \cite{Podolsky}). 

The towers of charges introduced in Sec. \ref{tc} are slightly simpler in the instance under consideration. Indeed, the Carrollian conformal Killings used in expressions \eqref{carkappaK-res} and \eqref{carkappaK-cot-res} are 
\eqref{comXclock} with 
\begin{equation}
\label{comclock}
C(t,\zeta,\bar \zeta)= t P(\zeta,\bar \zeta)
\end{equation}
(see \eqref{comclock}). Observe also that $ \tilde K^i =\tilde K_{\text{Cot}}^i =0$ so that $ \tilde{\mathcal{K}}_{(s)}=\tilde{\mathcal{K}}_{\text{Cot}\,(s)}=0$. Generically, 
 $ \mathcal{K}_{(s)}$ and  $\mathcal{K}_{\text{Cot}\,(s)}$ are non-zero though, because the conformal Killing vectors are not necessarily strong and due to the time dependence, here encoded exclusively  in their component $\xi^{\hat t}$.  The corresponding charges are ultimately expressed as integrals of combinations of $\hat k$, $\hat{\bar k}$, 
$\hat \tau$, $\hat{\bar \tau}$,  $\ast \varpi$, $\varpi$, and of their derivatives.\footnote{Although the components $b_i$ of the Ehresmann connection enter the expression of the Carrollian charges \eqref{carconch}, upon integration by parts, they are traded for $\ast \varpi$ or $\varpi$.}

For concreteness, we will illustrate the above with the distinctive strong Carrollian conformal Killing field $\partial_t$, i.e. the generator of the Ehlers--Geroch bulk three-dimensional reduction. For this Killing field, the ``tilde'' Carrollian charges vanish. In example, for the leading charges ($s=0$ in the coding of Sec.  \ref{tc}), we find\footnote{Using \eqref{carkappaK-res} and \eqref{carkappaK-cot-res}  with $\xi^{\hat{t}}=1$ and $\xi^i=0$, we find  $\kappa= -\varepsilon$, $K^i= -\frac{1}{8\pi G}\ast\!\chi^i$, $\kappa_{\text{Cot}} =-c$ and $K_{\text{Cot}} ^i=-\chi^i$.}
\begin{equation}
\label{Qelstatmass}
Q_{\text{em}}=\int_{\mathscr{S}}\frac{\text{d}\zeta \wedge \text{d}\bar \zeta}{\text{i}P^2}\left(8\pi G \varepsilon+\varpi K\right), 
\quad
Q_{\text{mm}}=\int_{\mathscr{S}}\frac{\text{d}\zeta \wedge \text{d}\bar \zeta}{\text{i}P^2}\left(-c+\ast \varpi K\right), 
\end{equation}
up to boundary terms with respect to \eqref{intcarconch} and \eqref{intcarconchcot} (and a factor $-8\pi G$ for the former),
handily combined into
\begin{equation}
\label{Qelstatmasscc}
Q_{\text{m}}=Q_{\text{mm}}+\text{i}\, Q_{\text{em}}=\int_{\mathscr{S}}\frac{\text{d}\zeta \wedge \text{d}\bar \zeta}{\text{i}P^2}\left(\hat \tau+\hat \varpi  K\right).
\end{equation}
The indices stand for magnetic and electric masses. These mass definitions carry some arbitrariness since, as a consequence of time independence, each of the terms in the integrals provide a separate well-defined charge. We will turn back to this when discussing the action of the M\"obius group, in Sec. \ref{behch}. 

Following Sec.  \ref{tc}, the subleading mass charges associated with the strong Carrollian conformal Killing field $\partial_t$ are captured in 
\begin{equation}
\label{Qelstatmassccn}
Q_{\text{m}\, (s)}=\int_{\mathscr{S}}\frac{\text{d}\zeta \wedge \text{d}\bar \zeta}{\text{i}P^2}
\left(\hat \tau+\hat \varpi  K\right)\ast\! \varpi^{2s}
\end{equation}
and define the higher-$s$ \emph{mass multipole moments}.
In the instance of the  $K=1$ Kerr--Taub--NUT family displayed in Eqs. \eqref{tau-TN}, \eqref{f-TN}, \eqref{bz-ckKTN}, \eqref{astvp-ckKTN}
with $A=\nicefrac{1}{2}$ and $D=1$,
we find:
\begin{equation}
\label{QelstatmassccnKTN}
Q_{\text{m}\, (s)}=4\pi \text{i}\left(M+\text{i}n\right)\left(\frac{(n+a)^{2s+1}-(n-a)^{2s+1}}{a(2s+1)}\right).
\end{equation}
For this set of solutions, $\upxi_1$ in \eqref{XK} is a strong Carrollian Killing vector, which brings its own Carrollian rotational charges. Again the ``tilde''  (Eqs. \eqref{tildeintcarconch} and \eqref{tildeintcarconchcot}) vanish whereas the ``non-tilde'' (see. \eqref{carkappaK-res} and \eqref{carkappaK-cot-res}) are combined in
the complex higher-$s$ \emph{angular-momentum multipole moments}
\begin{equation}
	  	Q_{\text{r}\,(s)}=\int_{\mathscr{S}}\frac{\text{d}\zeta \wedge \text{d}\bar \zeta}{\text{i}P^2}6\zeta\bar{\zeta}\left(
\frac{n+\text{i}M}{P^2}(a-nP)\left(n+a-\frac{2a}{P}\right)^{2s}-\frac{2a}{P^2} \left(n+a-\frac{2a}{P}\right)^{2s+1}	
		\right)
\label{QelstatanmomccnKTNgen}
\end{equation}
with $P=1+\frac{1}{2}\zeta\bar\zeta$,
which are non-zero if one rotation parameter $a$ or $n$ is present. We find for example: 
\begin{equation}
	  	Q_{\text{r}\,(0)}= -8\pi \left[a(n+\text{i}M)+3n(n-\text{i}M)\right].
\label{QelstatanmomccnKTN}
\end{equation}
Expressions \eqref{QelstatmassccnKTN} and \eqref{QelstatanmomccnKTN} are in line with the results obtained in Refs. \cite{  GerochMI, GerochMII, Hansen, Fodor} (see also \cite{FRODDEN2022137264}, where the electric part of $Q_{\text{r}\,(0)}$ is given)
using standard methods circumscribed to bulk dynamics. They provide conserved moments since the divergences \eqref{dives} and \eqref{divms} vanish. 

\section{Ehlers transformations}\label{RFCB}

\subsection{Bulk reduction and M\"obius action on the boundary}

Our next and pivotal task is to unravel the action of the Ehlers group \eqref{ehlers} on the boundary Carrollian observables, using the expression of the bulk Ricci-flat metric \eqref{exp-ric-fl} assumed to possess a time-like Killing vector field. We will focus in the present work on the restricted class of resummable metrics \eqref{exp-ric-fl-res}, as exploited in Sec. \ref{static}, i.e. equipped with a time-like Killing field $\upxi=\partial_t$ and $\Omega=1$.

In order to proceed, we are called to follow the steps for the Geroch reduction described in Sec.~\ref{EG}, i.e. determine $\tau$ as defined in \eqref{tau} for the metric  \eqref{exp-ric-fl-res} with $\lambda$ and $\omega $ given in \eqref{lambda}, \eqref{doublew} and \eqref{omega}. These should be expanded in inverse powers of $r$ and thus deliver the boundary ingredients together with their transformations following  \eqref{ehlers}. A remark should be made before hand. The Geroch reduction is followed by an oxidation, which defines the novel Ricci-flat solution. Nothing guarantees in this course that the oxidized metric will assume again the form  \eqref{exp-ric-fl-res}. Actually it doesn't and a redefinition of the radial coordinate is necessary to bring it back into the expected original gauge. 

It is convenient for the present mission to adopt the Cartan frame defined in \eqref{bulkfrcofr}, leading to the bulk metric 
\begin{equation}
\label{resmetexp-stat}
\begin{array}{rcl}
&g_{\hat t \hat t}=  \frac{1 }{\rho^2}\left(8\pi G\varepsilon r+\ast \varpi c\right)-K
,\quad
g_{\hat ti}=  \ast\partial_{i}\ast\!\varpi ,\quad
g_{\hat tr}=-1, & \\
&g_{ri}= 0,\quad
g_{rr}=  0,\quad
g_{ij}= \rho^2 a_{ij} &
\end{array}
\end{equation}
obtained using \eqref{resmetexp}, assuming $t$-independence and $\Omega=1$. In this expression $\ast \varpi$,  $K$ and $c$ are given in Eqs. \eqref{astvpitinv}, 
 \eqref{Ktinv} and
 \eqref{c-Carrolholtinv}.
The Killing form reads:
\begin{equation}
\label{Kmu}
\upxi=\left(K- \frac{1 }{\rho^2}\left(8\pi G\varepsilon r+\ast \varpi c\right)\right)\upmu
+\ast\partial_{i}\ast\!\varpi \text{d}x^i
- \text{d}r,
\end{equation}
with norm 
\begin{equation}
\lambda=\frac{8\pi G\varepsilon r+\ast \varpi c}{\rho^2}-K
.
\end{equation}
For the twist we use Eq. \eqref{doublew}, expressed as
\begin{equation}
\label{wstat}
\text{w}= - \star\! \left(\upxi\wedge \text{d}\upxi\right),
\end{equation}
where ``$\star$'' stands for the four-dimensional Hodge duality. The latter one-form is exact on-shell and we find the following potential (Eq. \eqref{omega}):
\begin{equation}
\omega =\frac{8\pi G \varepsilon \ast\! \varpi-cr}{\rho^2} +K^\ast.
\label{omega-KTN}
\end{equation}
On-shellness is implemented here through boundary dynamics as summarized in Sec. \ref{static}, i.e. in Eqs. \eqref{charm}, \eqref{k} and \eqref{kst}.

Inserting the above results into Eqs. \eqref{tau} and using \eqref{taudef}, we find 
\begin{equation}
\label{tau-tau-bar}
\tau =
\frac{\hat \tau}{r+\text{i}\ast \!\varpi } -\text{i}\hat k 
.
\end{equation}
Likewise, we obtain the Geroch reduced and rescaled metric \eqref{met-h-tilde}: 
\begin{equation}
\label{redmetexp-stat}
\tilde h_{AB}\text{d}x^A\text{d}x^B=-\left(\text{d}r- \ast\partial_{k}\ast\!\varpi \,\text{d}x^k \right)^2+ \lambda  \rho^2 a_{ij}\text{d}x^i \text{d}x^i.
\end{equation}
With this, $\tau$ given in \eqref{tau-tau-bar} unsurprisingly solves the reduced Einstein's equations \eqref{eq: funda equa}.

The premier Ehlers transformation rules are \eqref{ehlers} and the invariance of $\tilde h_{AB}$. From these follows the rest of the construction, i.e. the transformation of $h_{AB}$ and the oxidation toward $g_{AB}'$. In the present framework, we have to some extent locked the gauge, via the resummed bulk expression \eqref{exp-ric-fl-res}. Ehlers transformations are not designed a priori to maintain this form, and they are generally expected to require further coordinate transformations. It is rather remarkable that,
to this end,  a local (i.e. celestial-sphere dependent) shift in the radial coordinate suffices.

Using for convenience holomorphic and antiholomorphic coordinates as introduced in App. \ref{carcot3}, expression  \eqref{redmetexp-stat} is recast as follows:
\begin{equation}
\label{redmetexp-stat-hom}
\tilde h_{AB}\text{d}x^A\text{d}x^B=-\left(\text{d}r- \text{i}\partial_{\zeta}\ast\!\varpi \,\text{d}\zeta+\text{i} \partial_{\bar \zeta}\ast\!\varpi \,\text{d}\bar \zeta  \right)^2+ \frac{(\tau-\bar\tau)  (r+\text{i}\ast \!\varpi )(r-\text{i}\ast \!\varpi )}{\text{i} P^2}\text{d}\zeta\text{d}\bar\zeta.
\end{equation}
Combining \eqref{ehlers} with \eqref{tau-tau-bar}, we obtain the following \emph{boundary} transformations:
\begin{eqnarray}
\label{btautr}
\hat \tau'&=& -\frac{\hat \tau}{\left(\gamma \hat k +\text{i}\delta\right)^2},
\\
\label{bktr}
\hat k'&=&\text{i} \frac{\alpha \hat k +\text{i}\beta}{\gamma \hat k +\text{i}\delta},
\\
\label{bvarpitr} 
 \hat\varpi'&=& \hat\varpi 
 + \frac{\gamma \hat \tau}{\gamma \hat k +\text{i}\delta} 
 \end{eqnarray}
and 
\begin{equation}
\label{bPtr}
P'=\frac{P}{\left\vert\gamma \hat k +\text{i}\delta\right\vert},
\end{equation}
plus the radial shift\footnote{One could alternatively adopt a new radial coordinate defined as $\tilde r= r + \varpi$ that is invariant under M\"obius transformations. This is actually mandatory in order to reach boundary $SL(2,\mathbb{R})$-covariant tensors from the bulk, as we will discuss in Sec. \ref{behch}. It furthermore coincides with the radial coordinate of Ref.  \cite{Stephani:624239} \S 29 provided $r_0 =-\varpi$ (origin of the affine parameter along the geodesic congruence tangent to $\partial_r$ -- see footnote \ref{origin}).
\label{rtilde}}
\begin{equation}
\label{brtr}
r'= r+  \frac{\text{i}}{2}\left( \frac{\gamma \hat \tau}{\gamma \hat k +\text{i}\delta} -  \frac{\gamma \hat{ \bar\tau}}{\gamma \hat{\bar k} -\text{i}\delta}\right).
\end{equation}
These transformation rules  leave indeed \eqref{redmetexp-stat-hom} invariant. As advertised earlier, they  are \emph{local}, providing a direct transformation \eqref{bPtr} of the boundary metric. The transformation of the energy density $\varepsilon$ is obtained from \eqref{btautr} using  \eqref{taudef}:
\begin{equation}
\label{Trepsilon}
8\pi G\varepsilon'=\frac{8 \pi G \varepsilon\left(\left(\gamma K^\ast+\delta\right)^2-\gamma^2 K^2 \right)-2c
 \gamma K\left(\gamma K^\ast+\delta\right)}{\left(\gamma^2 K^2 +\left(\gamma K^\ast+\delta\right)^2\right)^2}
	\,.
\end{equation}
The transformation of $c$ is inferred similarly:
\begin{equation}
\label{Trcotton}
	c'=\frac{c\left(\left(\gamma K^\ast+\delta\right)^2 -\gamma^2 K^2\right)+16 \pi G \varepsilon \gamma K\left(\gamma K^\ast+\delta\right)}{\left(\gamma^2 K^2 +\left(\gamma K^\ast+\delta\right)^2\right)^2}\,.
\end{equation}
All these rules are compatible with Eqs. \eqref{Ktinv} and \eqref{c-Carrolholtinv}. Finally the 
transformations of the Carrollian Cotton tensors are reached using the above results  combined with Eqs. \eqref{carcotch}, \eqref{carcotps} and \eqref{carcotPs}.

The transformation of the Ehresmann connection is obtained directly from the expressions reached for the latter in \eqref{bz-nonck},
\eqref{bz-ck} and \eqref{bz-K0}.	 To this end, observe that in the constant-$\hat k$ instance, $A$, $B$, $\bar B$ and $D$ transform with a factor $\nicefrac{1}{\left\vert\gamma \hat k +\text{i}\delta\right\vert}$ in order to comply with \eqref{bPtr}.  Similarly, $f(\zeta)$ and $Z(\zeta)$, introduced in \eqref{astvp-ck} and \eqref{astvp-K0}, must be respectively invariant and transforming as
\begin{equation}
\label{Z-tr}
Z'(\zeta)=Z(\zeta)+\text{i}  \frac{\gamma \hat \tau(\zeta)}{\gamma \hat k +\text{i}\delta} ,
\end{equation}
so that \eqref{bvarpitr} be fulfilled. 

Let us mention for completeness that once the M\"obius transformation is performed on the boundary, the reconstruction of the new Ricci-flat solution is straightforward using the boundary-to-bulk formula \eqref{exp-ric-fl-res}, expressed with primed data -- except for the unaltered boundary coordinates $\left\{t, \zeta, \bar \zeta\right\}$. This is equivalent to the 
oxidation procedure operated from three to four dimensions along the lines of  Eqs.~\eqref{F}, \eqref{eta}, \eqref{eta-norm} and \eqref{newkil} with  
\begin{equation}
\label{etaprime}
\upeta'=-\upmu'
-\frac{1}{\lambda'}\left(\text{d}r'- \text{i}\partial_{\zeta}\ast\!\varpi' \,\text{d}\zeta+\text{i} \partial_{\bar \zeta}\ast\!\varpi' \,\text{d}\bar \zeta\right), \quad  \upmu' = -\text{d}t +b_{\zeta}'\text{d}\zeta+b_{\bar \zeta}'\text{d}\bar \zeta,
\end{equation}
finally leading to \eqref{newmet}, which assumes the form  \eqref{exp-ric-fl-res} primed. The new bulk Killing vector $\upxi'=\lambda'\upeta'$ is again $\partial_t$.

In the example of the Kerr--Taub--NUT family treated at the end of Sec. \ref{static}, the specific choices of $P=\frac{1}{2}\zeta\bar\zeta +1$, $K=1$ and  $K^\ast=0$ (this was not explicitly demanded) are stable only under  $\left(\begin{smallmatrix}      \cos \chi & \sin \chi \\ -\sin \chi & \cos \chi  \end{smallmatrix}\right) \in SL(2,\mathbb{R})$. For this transformation, using \eqref{btautr} we find $M'+\text{i}n'=(M+\text{i}n)\text{e}^{-2\text{i}\chi}$. Observe that \eqref{bvarpitr} will switch on a non-zero $\varpi$ though, as opposed to its original value in the family at hand (see footnote \ref{tau0}).

\boldmath
\subsection{Charges and $SL(2,\mathbb{R})$ multiplets} \label{behch}
  \unboldmath

Carrollian charges have been introduced in Sec. \ref{tc} and further discussed for stationary and algebraic spacetimes in Sec. \ref{static}. Two generic charges were found and displayed in \eqref{Qelstat} and \eqref{Qelstatmasscc}. The former is purely geometric and stands for the integrated curvature of the celestial sphere; the latter carries genuine dynamic information captured in the electric and magnetic masses.  It is legitimate to wonder how these quantities behave under M\"obius transformations, and possibly tame them in $SL(2,\mathbb{R})$ multiplets. Although ideally this programme should be conducted for reductions along generic bulk Killing fields and no special algebraic structure -- these would be non-resummable, i.e. of the form \eqref{exp-ric-fl}, and labelled by a possibly plethoric set of independent charges --  we will pursue it here for illustrative purposes  in the restricted framework at hand. 

The curvature charge $Q_{\text{ec}}$ in \eqref{Qelstat} is invariant under Ehlers' $SL(2,\mathbb{R})$, and this is inferred using the transformation laws \eqref{bktr} and \eqref{bPtr}. The mass charge $Q_{\text{m}}$, Eq.  \eqref{Qelstatmasscc},  is not, but its transformation (see  \eqref{btautr},  \eqref{bktr} and  \eqref{bvarpitr}) suggests that it might belong to some $SL(2,\mathbb{R})$ multiplet or, more accurately, that it may be modified to this end -- we have this freedom owing to time independence. Actually, a slight amendment to the charge $Q_{\text{m}}$, namely
\begin{equation}
\label{Qelstatmasscct}
Q_{\text{m}}'=\int_{\mathscr{S}}\frac{\text{d}\zeta \wedge \text{d}\bar \zeta}{\text{i}P^2}\left(\hat \tau+2\hat \varpi  K\right),
\end{equation}
is $SL(2,\mathbb{R})$-invariant. We can even go further and apply the following pattern
to generate $SL(2,\mathbb{R})$ triplets. Suppose we identify a Carrollian two-form $\pmb{v}$ transforming under  $SL(2,\mathbb{R})$ as 
\begin{equation}
\label{gentr}
 \pmb{v}\to  \pmb{v}'=- \pmb{v} \left(\gamma \hat{\bar k} -\text{i}\delta\right)^2.
\end{equation}
This allows to design an $SL(2,\mathbb{R})$ two-form triplet, i.e. a symmetric rank-two tensor, transforming  as
\begin{equation}
\label{gentrtriplet}
\begin{pmatrix}      \pmb{v}_3' & \pmb{v}_2' \\ \pmb{v}_2' & \pmb{v}_1'  \end{pmatrix}
=\begin{pmatrix}      \alpha & \beta \\ \gamma & \delta  \end{pmatrix}
\begin{pmatrix}      \pmb{v}_3 & \pmb{v}_2 \\ \pmb{v}_2 & \pmb{v}_1  \end{pmatrix}
\begin{pmatrix}      \alpha & \gamma \\ \beta & \delta  \end{pmatrix},
\end{equation}
where 
\begin{equation}
\label{gentriplet}
 \pmb{v}_1= \pmb{v}, \quad
 \pmb{v}_2=\text{i} \hat{\bar k} \pmb{v}, \quad
 \pmb{v}_3=-\hat{\bar k}^2  \pmb{v}.
\end{equation}
The same holds for the complex-conjugate triplet: 
 $\bar{\pmb{v}}_1= \bar{\pmb{v}}$, 
$ \bar {\pmb{v}}_2=-\text{i} \hat{ k} \bar{\pmb{v}}$ 
and $ \bar{\pmb{v}}_3=-\hat{k}^2  \bar{\pmb{v}}$. An $SL(2,\mathbb{R})$ triplet of charges is thus reached as
\begin{equation}
\label{QI}
Q_I=\int_{\mathscr{S}}\pmb{v}_I, \quad I=1,2,3,
\end{equation}
and $Q\equiv Q_1 Q_3 -Q_2^2$ is invariant under M\"obius transformations.

The above strategy can be readily applied. Two-forms transforming as in \eqref{gentr} can be found, inspired by the structures of  the charge \eqref{Qelstatmasscc} and of the Carrollian currents \eqref{carkappaK-res} and \eqref{carkappaK-cot-res}, given the expressions of the Carrollian twist 
 \eqref{astdvpitinvc}, the Carrollian  curvature
 \eqref{Ktinv}, and the Carrollian Cotton tensors \eqref{c-Carrolholtinv}, 
 \eqref{carcotch}, 
 \eqref{carcotps} and 
 \eqref{carcotPs}. We here exhibit two such Carrollian forms:
\begin{eqnarray}
\label{x-form}
\pmb{x}&=& -\frac{\hat \tau}{2(\hat k+\hat{\bar k})}\frac{\text{d}\zeta \wedge \text{d}\bar \zeta}{\text{i}P^2},
\\
\label{y-form}
\pmb{y}&=& -\left(\frac{P}{\hat k+\hat{\bar k}}\right)^2\partial_\zeta\hat k \, \partial_{\bar\zeta}\hat\varpi
\frac{\text{d}\zeta \wedge \text{d}\bar \zeta}{\text{i}P^2}.
\end{eqnarray}
These lead  along \eqref{QI} to two triplets of charges, which do not carry more information than the original \eqref{Qelstat}  and \eqref{Qelstatmasscc} though -- in the constant-$\hat k$ paradigm, which is in fact the most generic, these are $K$, $M$, $n$ and possibly $a$, and the second triplet vanishes.

The last item in our Carrollian agenda is to setting the relationship amongst the charges introduced here using purely boundary methods and those computed directly by standard bulk techniques. This sort of question definitely deserves to be addressed  in more general situations than ours, i.e. in the presence of a large set of non-trivial surface charges computed e.g. within covariant phase-space formalism \cite{BB}. Nonetheless some relevant observations can be made here, in relation with the original discussion on charges of Ref. \cite{Geroch}, in which the above two-forms \eqref{x-form} and \eqref{y-form} turn out to play a prominent role.
 
In  Ref. \cite{Geroch}, an  $SL(2,\mathbb{R})$ triplet of bulk two-forms, leading to surface charges upon integration on the celestial sphere of $\mathcal{M}$, is obtained by oxidizing the following two-form triplet of  $\mathcal{S}\equiv\nicefrac{\mathcal{M}}{\text{orb}(\upxi)}$ (Eqs. (18) and (16) of the quoted reference):
\begin{equation}
\label{reductripl}
\begin{array}{rcl}
\displaystyle{\text{V}}_1&=&\displaystyle{ \frac{1}{(\tau - \bar{\tau})^{2}}\star^3_{\tilde{\text h}}\left(\text{d}\tau +\text{d} \bar{\tau}\right),} \crbig
\displaystyle{\text{V}}_2&=&\displaystyle{ \frac{1}{(\tau - \bar{\tau})^{2}}\star^3_{\tilde{\text h}}\left(\bar{\tau} \text{d}\tau + \tau \text{d} \bar{\tau}\right),}  \crbig
\displaystyle{\text{V}}_3&=&\displaystyle{ \frac{1}{(\tau - \bar{\tau})^{2}}\star^3_{\tilde{\text h}}\left(\bar{\tau}^2\text{d}\tau + \tau^2  \text{d}\bar{\tau}\right),}  
\end{array}
\end{equation}
where ``$\star^3_{\tilde{\text h}}$'' stands for the three-dimensional Hodge-dual on $\mathcal{S}$ equipped with $\tilde{h}_{AB}$ displayed in \eqref{redmetexp-stat-hom}. It is remarkable that the asymptotic limit of this two-form triplet coincides with those designed earlier from 
Carrollian boundary considerations. This statement is captured in the following result:
\begin{equation}
\label{reductripllim}
\lim_{\tilde r \to\infty} \begin{pmatrix}    \text{V}_3 &  \text{V}_2 \\ \text{V}_2 &\text{V}_1  \end{pmatrix}
=  \begin{pmatrix} -\hat{\bar k}^2  (\pmb{x}+ \pmb{y})-\hat{ k}^2(\bar{\pmb{x}}+\bar{\pmb{y}}) & \text{i} \hat{\bar k}(\pmb{x}+ \pmb{y})-\text{i} \hat{ k}(\bar{\pmb{x}}+\bar{\pmb{y}}) \\ \text{i} \hat{\bar k}(\pmb{x}+ \pmb{y})-\text{i} \hat{k}(\bar{\pmb{x}}+\bar{\pmb{y}}) &\pmb{x}+ \pmb{y}+\bar{\pmb{x}}+\bar{\pmb{y}}  \end{pmatrix}
,
\end{equation}
where $\tilde r= r+\varpi$ was introduced  in footnote \ref{rtilde} as an $SL(2,\mathbb{R})$-invariant radial coordinate, which must be used here in order to guarantee that the limit preserves the $SL(2,\mathbb{R})$ behaviour.
 
\section{Conclusions}

When a four-dimensional spacetime geometry is invariant under the action of a one-dimensional group of motions, a reduction can be performed and vacuum Einstein dynamics reveals a symmetry under M\"obius transformations. Our main motivation was to exhibit this action from a holographic perspective, namely on the three-dimensional boundary of the Ricci-flat configuration at hand. We have successfully reached this goal for a class of \emph{resummable} or \emph{integrable} metrics, which are algebraic in Petrov' classification and possess a time-like isometry. All of our findings can be extended to embody any Ricci-flat spacetime possessing an isometry at the expense of an augmented 
technical difficulty due to (\romannumeral1) the use of generic Killing vectors with Ehlers action ending outside the class of algebraically special, resummable metrics \eqref{exp-ric-fl-res},\footnote{See e.g. \cite{Contopoulos:2015wra}, where examples of space-like Killings are displayed with Ehlers groups connecting Petrov special to Petrov general Ricci-flat spacetimes (more recent works in a similar spirit are Refs. \cite{Astorino:2023elf,Barrientos:2023tqb
}), and \cite{Mars} for a mathematical essay on the behaviour of the Weyl tensor under Ehlers' M\"obius group.} and (\romannumeral2)
the presence of an indefinitely increasing number of independent boundary observables transforming under $SL(2,\mathbb{R})$. The main features of the boundary $SL(2,\mathbb{R})$ action are however clearly captured by the simplest case treated here and we will now summarize them.

At the heart of the boundary M\"obius transformations one finds the Carrollian Cotton tensors. The latter are a set of descendants of the original boundary pseudo-Riemannian Cotton, reached in the zero-speed-of-light limit. One finds in particular a scalar $c$, which is a
dual-mass aspect, naturally combined with the Bondi mass aspect, another Carrollian scalar identified with the boundary Carrollian fluid energy density $\varepsilon$. The M\"obius transformation hence mixes the geometric boundary variables i.e. those which determine the boundary itself with the dynamical variables like the boundary fluid (this is one of the infinite data, made redundant in the resummable situation studied here).  Our analysis reveals that this duality transformation on the boundary is \emph{algebraic} i.e.  \emph{local} for the metric, Ehresmann connection, field of observers, and for every other Carrollian boundary data. This is an important achievement summarized in Eqs. \eqref{btautr},   \eqref{bktr},   \eqref{bvarpitr}
 and   \eqref{bPtr}, rooted in the decoupling of $r$ close to the boundary.

An aside message this analysis conveys is the role of the Cotton tensor, which is manifestly dual to the energy/momentum. Before the advent of flat holography and Carrollian physics, the boundary Cotton tensor had been recognized in AdS/CFT as an unavoidable boundary trait carrying information on the bulk magnetic charges such as the nut  \cite{Mukhopadhyay:2013gja, Gath:2015nxa} (see also \cite{deFreitas:2014lia}).  The M\"obius transformation   \eqref{btautr} for the $SO(2)\subset SL(2,\mathbb{R})$ subgroup 
 was actually anticipated as a relationship on the conformal pseudo-Riemannian boundary of four-dimensional  Einstein spacetimes \cite{Petropoulos:2015fba, Petkou:2015fvh}, in an attempt to relate electric and magnetic solutions to Einstein's equations. Although such dual solutions exist irrespective of the cosmological constant, the relevant subgroup of Ehlers' breaks down for $\Lambda\neq 0$ \cite{Leigh:2014dja}. The bulk duality relationship fades in this case, but persists asymptotically and reveals on the conformal boundary. What we find here is a $\Lambda$-to-zero limit of this relationship. 

Notwithstanding their role in boundary Ehlers duality manifestation, the Carrollian Cotton tensors obey off-shell conservation properties and  generate towers of magnetic charges, some of them being conserved. This property is not exclusive to Ricci-flat spacetimes and Carrollian boundaries. Einstein bulk spacetimes and pseudo-Riemannian boundaries do provide a conserved Cotton tensor, which contracted with any boundary conformal Killing vector leads to a conserved current, hence a conserved charge. This powerful tool is undermined by the limited -- if any --  number of conformal isometries on arbitrary three-dimensional Riemannian spacetimes. The remarkable spin-off  about Carrollian boundaries is the existence of an infinite-dimensional conformal group, which makes this method of charge determination a serious alternative to the more standard bulk asymptotic techniques. 
Following the Cotton pattern, electric towers of charges are constructed with the fluid dynamical data, which can only enjoy on-shell conservation -- the same would hold in AdS boundaries with the aforementioned  limitation. On both electric and magnetic sides, the towers of charges are multiplied ad nauseam, beyond their leading components. 

Our present investigation on towers of charges designed from a boundary standpoint is radically novel and deserves a systematic extension. It has been here confined in the integrable case, where the infinite set of observables is redundant and shrinks to the elementary ``leading'' data -- our tentative definition of subleading currents might have turned too naive, hadn't it reproduced successfully the multipole moments. Moreover, our main goal being primarily on boundary Ehlers action, we have assumed a time-like bulk isometry, which further reduces this set. Besides, the chosen time-like Killing field was aligned with the fibre of the boundary Carrollian structure, which screens the black-hole acceleration parameter and avoids exploring head-on the uncharted subject of Carrollian reductions. The latter is the mathematical tool to be developed for unravelling the bulk-to-boundary relationship of hidden symmetries in Ricci-flat spacetimes. It could encompass bulk reductions along space-like isometries, which are interesting because they leave room for gravitational radiation,\footnote{The Petrov-algebraic spacetimes \eqref{exp-ric-fl-res} accommodate  axisymmetric time-dependent solutions of the Robinson--Trautman type, whose final state is the C-metric -- see. \cite{Stephani:624239} \S 28.1.} probing the interplay between Ehlers M\"obius group, time evolution and charge non-conservation. Last, we did not address the question of the charge algebra and its potential central extensions, or discussed other more general related physical aspects. All this calls for a thorough comparison to alternative approaches such as those of Refs. \cite{Godazgar:2018vmm, Godazgar:2018qpq, Godazgar:2018dvh, Kol:2019nkc,Godazgar:2020gqd,Godazgar:2020kqd,Oliveri:2020xls,Kol:2020vet} based on Newman--Penrose formalism -- or to applications \cite{Compere:2017wrj, Seraj:2021rxd,Seraj:2022qyt,Godazgar:2022jxm, Awad:2022jgn,Godazgar:2022pbx, Grant:2021hga}.

In the chapter of charges, in spite of the various limitations just stated, we have successfully described the Ehlers M\" obius action, and discussed the organisation of available charges in $SL(2,\mathbb{R})$ multiplets. This enabled us to recover from a Carrollian viewpoint the triplet of Komar charges inferred by Geroch in its original publication \cite{Geroch}. Again, this result should be considered as a first step toward a  methodical $SL(2,\mathbb{R})$ taming of the above towers of electric/magnetic currents and charges in more general situations. These objects should include the boundary attributes of the bulk Weyl tensor, whose behaviour under M\"obius transformations has been addressed in \cite{Mars}.

The importance of the boundary covariantization -- Carroll and Weyl -- is yet another feature we would like to stress, as it hasn't been sufficiently appreciated in the literature. This characteristic is absent from Bondi or Newman--Unti gauges, where the formalism might suggest that the relevant part of the conformal boundary is its two-dimensional spatial section -- the celestial sphere. We heavily insist on the three-dimensional and Carrollian nature of the boundary, which is made manifest in the gauge we have been using. In ordinary AdS/CFT holography the Fefferman--Graham gauge is superior for this reason. One should likewise use a truly boundary-covariant gauge in flat holography and take advantage of it, as we modestly did for exhibiting the action of Ehlers' group, or for discussing the charges and their conservation. No boundary approach of this sort would have been possible in the more conventional  gauges. Correspondingly, flat holography based on a purely celestial gauge is bound to be incomplete.

It is worth mentioning that Ehlers' $SL(2,\mathbb{R})$ group is the first and simplest example of a hidden symmetry. As pointed out in the introduction (see the references proposed there), more involved reductions reveal richer symmetries and the underlying dynamics is captured by elegant sigma models in various dimensions. Recasting this knowledge in a holographic fashion, we could possibly learn more, or at least differently, not only about hidden symmetries but also on flat holography. Carrollian reductions might  again be the appropriate tool. 

On a more speculative tone, our results suggest that a boundary analysis might reveal more general or unexpected duality properties.
The paradigm of anti-de Sitter spacetimes, where the $SL(2,\mathbb{R})$ is broken in the bulk but restored on the boundary, calls for a  systematic investigation that would complement the heuristic discussion of Ref. \cite{Petropoulos:2015fba}, and possibly uncover novel instances of boundary duality symmetries, associated e.g. with an asymptotic Killing field rather than a plain reduction along Killing orbits. One could even be more audacious and entertain the idea of a ``boundary'' analysis for half-flat spaces (this is vaguely motivated by footnote \ref{selfduality}), which have attracted some attention in relation with $w_{1+\infty}$ symmetry (see the original works  \cite{Penrose:1976twist,Plebanski:1985,Park:1990 } and \cite{Strominger:walgebras,Adamo:2021lrv, Compere:2022zdz, Mason:2022hly, Freidel:2021ytz} for a recent emanation). The main caveat foreseen here is the absence of Carrollian boundaries in Euclidean gravity, but this could be evaded in the ultra-hyperbolic instance ($2+2$ signature).

 \section*{Acknowledgements}

We would like to thank our colleagues G. Bossard, L. Ciambelli, A. Fiorucci, M. Godazgar, N. Lambert, C.~Marteau, B. Oblak, G. Papadopoulos, A. Petkou, R. Ruzziconi, A.~Seraj, K. Siampos, A. Stergiou and P.~West for useful discussions and feedback. Nehal Mittal acknowledges the CPHT of the Ecole Polytechnique for hospitality during his M1 internship in 2021 and 
\textsl{DIM Quantip} for funding his PhD fellowship at the LKB. The work of D. Rivera-Betancour was funded by Becas Chile (ANID) Scholarship No.~72200301. 
The work of M. Vilatte  was supported by the Hellenic Foundation for Research and Innovation (H.F.R.I.) under the \textsl{First Call for H.F.R.I. Research Projects to support Faculty members and Researchers and the procurement of high-cost research equipment grant} (MIS 1524, Project Number: 96048).

\appendix

\section{Carrollian covariance in arbitrary dimension} \label{carman}

Carroll structures on $\mathscr{M}= \mathbb{R} \times \mathscr{S}$   were introduced in Sec. \ref{brrm} with emphasis on the covariance properties they enjoy when the time coordinate is aligned with the fiber of the structure. In the present appendix we will  elaborate on this subject, treating in particular Carrollian covariant and Weyl-covariant derivatives. 

The Carrollian transformations (Eqs. \eqref{cardifs} and \eqref{carj}) are connection-like (non-covariant) for $\partial_i$ and $b_i$, and density-like for $\partial_t$ and $\Omega$: 
\begin{equation}
\partial^\prime_j=J^{-1i}_{\hphantom{-1}j}\left(\partial_i-
\frac{j_i}{J}\partial_t\right),\quad b^{\prime}_{k}=\left( b_i+\frac{\Omega}{J} j_i\right)J^{-1i}_{\hphantom{-1}k},\quad 
\partial^\prime_t=\frac{1}{J}\partial_t, \quad
\Omega^{\prime }=\frac{\Omega}{J}.
\end{equation}
The vector fields dual to the forms $\text{d}x^i$ are 
\begin{equation}
\label{dhat}
\hat\partial_i=\partial_i+\frac{b_i}{\Omega}\partial_t
\end{equation}
and transform covariantly under \eqref{cardifs} together with the metric \eqref{cardegmet}, and the fields \eqref{kert} and \eqref{kertdual}:
\begin{equation}
\upupsilon^{\prime}
=
\upupsilon,\quad
\upmu^{\prime}
=
\upmu,\quad
\hat\partial_i^\prime =
J^{-1j}_{\hphantom{-1}i} \hat\partial_j, \quad
a^{ ij\prime}=J^i_k J_l^ja^{kl}.
\end{equation}
The vectors $\hat\partial_i$ and $\upupsilon$ do not commute. They  define the \emph{Carrollian vorticity and acceleration}:
\begin{equation}
\label{carconcomderf}
\left[\hat\partial_i,\hat\partial_j\right]=
\frac{2}{\Omega}\varpi_{ij}\partial_t,\quad
\left[\upupsilon,\hat\partial_i\right]=
\frac{1}{\Omega}\varphi_i\partial_t,\quad\varpi_{ij}=\partial_{[i}b_{j]}+b_{[i}\varphi_{j]},\quad 
\varphi_i=\dfrac{1}{\Omega}\left(\partial_t b_i+\partial_i \Omega\right),
\end{equation}
similarly appearing in 
\begin{equation}
\label{dualcarconcomderf}
\text{d}\upmu=\varpi_{ij}
\text{d}x^i
\wedge
\text{d}x^j
+
\varphi_i
\text{d}x^i
\wedge
\upmu.
\end{equation}

A Carroll structure is also equipped with a metric-compatible and field-of-observers-compatible  connection (strong definition). Due to the degeneration of the metric, such a connection is not unique, but it can be chosen as the connection inherited from the parent relativistic spacetime (see footnote \ref{relpar}), 
\begin{equation}
\label{dgammaCar}
\hat\gamma^i_{jk}=\dfrac{a^{il}}{2}\left(
\hat\partial_j
a_{lk}+\hat\partial_k  a_{lj}-
\hat\partial_l a_{jk}\right),
\end{equation}
obeying $\hat\gamma^k_{[ij]}=0$,  $\hat\nabla_ia_{jk}=0$ and
leading to the Levi--Civita--Carroll spatial covariant derivative $\hat \nabla_i$.\footnote{Details on the transformation rules can be found in the appendix A.2 of Ref. \cite{CMPPS1}.}

The ordinary time-derivative operator $\frac{1}{\Omega}\partial_t$ acts covariantly on Carrollian tensors. However,  it is not metric-compatible because $a_{ij}$ depend on time and 
a  temporal covariant derivative is defined requiring  $\frac{1}{\Omega^\prime}{\hat D}^\prime_t=\frac{1}{\Omega}\hat D_t$ and $\hat D_ta_{jk}=0$. To this end,
we introduce a temporal connection (a sort of extrinsic curvature of the spatial section $\mathscr{S}$)
\begin{equation}
\label{dgammaCartime}
\hat\gamma_{ij}=\frac{1}{2\Omega}\partial_t a_{ij}
=\xi_{ij} + \frac{1}{d}a_{ij}\theta,\quad \theta=
\dfrac{1}{\Omega}              
\partial_t \ln\sqrt{a},
\end{equation}
which is a  symmetric Carrollian tensor spliting into the \emph{geometric Carrollian shear (traceless) and the Carrollian expansion }(trace). The action of  $\hat D_t$ on any tensor is obtained using Leibniz rule plus the action on scalars and vectors:
\begin{equation}
\label{Cartimecovdervecform}
\frac{1}{\Omega}\hat D_tV^i=\frac{1}{\Omega} \partial_tV^i+\hat\gamma^i_{\hphantom{i}j} V^j,\quad
\frac{1}{\Omega}\hat D_tV_i=\frac{1}{\Omega} \partial_tV_i-\hat\gamma_i^{\hphantom{i}j} V_j.
\end{equation}

The commutators of Carrollian covariant derivatives define Carrollian curvature tensors:
\begin{equation}
\begin{array}{rcl}
\left[\hat\nabla_k,\hat\nabla_l\right]V^i&=&\left(
\hat\partial_k\hat\gamma^i_{lj}
-\hat\partial_l\hat\gamma^i_{kj}
+\hat\gamma^i_{km}\hat\gamma^m_{lj}
-\hat\gamma^i_{lm}\hat\gamma^m_{kj}
\right)V^j+\left[\hat\partial_k,\hat\partial_l\right]V^i
\\
&=& \hat r^i_{\hphantom{i}jkl}V^j+
\varpi_{kl}\frac{2}{\Omega}\partial_{t}V^i,
\end{array}
\label{carriemann}
\end{equation}
where $ \hat r^i_{\hphantom{i}jkl}$ is the \emph{Riemann--Carroll} tensor. The Ricci--Carroll tensor and the Carroll scalar curvature are thus
\begin{equation}
\label{carricci-scalar}
\hat r_{ij}=\hat r^k_{\hphantom{k}ikj}\neq \hat r_{ji},\quad \hat r=a^{ij}\hat r_{ij}.
\end{equation}
Similarly, space and time derivatives do not commute:
\begin{equation}
\left[\frac{1}{\Omega}\hat D_t,\hat\nabla_i\right]V^j= 
\varphi_{i}\left(\left(\frac{1}{\Omega}\hat D_t
+
\theta
\right)V^j - \hat\gamma^{j}_{\hphantom{j}k}V^k\right)
-\hat\gamma_{i}^{\hphantom{i}k}\hat\nabla_k V^j
-d \hat r^j_{\hphantom{j}ik}V^k
\label{3carriemanntimetilde}
\end{equation}
with 
\begin{equation}
\hat r^j_{\hphantom{j}ik}=\frac{1}{d}
\left(\theta \varphi_i \delta^j_k 
+\hat \nabla_i \hat\gamma^{j}_{\hphantom{j}k}
-\frac{1}{\Omega} \partial_t \hat \gamma^j_{ik}
\right),
\quad
\hat r^j_{\hphantom{j}jk}=\hat r_k=\frac{1}{d}\left(\hat\nabla_j\hat\gamma^{j}_{\hphantom{j}k}
 -\hat\partial_k\theta
\right),
\label{carriemanntime}
\end{equation}
further Carrollian curvature tensors.

The boundary geometry -- be it pseudo-Riemannian or Carrollian -- enjoy conformal properties. 
Weyl transformations are defined through their action on  elementary geometric data
\begin{equation}
\label{weyl-geometry-abs}
a_{ij}\to \frac{1}{\mathcal{B}^2}a_{ij},\quad b_{i}\to \frac{1}{\mathcal{B}}b_{i},\quad \Omega\to \frac{1}{\mathcal{B}}\Omega
\end{equation} 
with $\mathcal{B}=\mathcal{B}(t,\mathbf{x})$ an arbitrary function. A Weyl-covariant derivative requires an appropriate connection built on $\varphi_i$ and  $\theta$ defined in \eqref{carconcomderf}
and
\eqref{dgammaCartime}, 
which transform as 
 \begin{equation}
 \label{weyl-geometry-2-abs}
 \varphi_{i}\to \varphi_{i}-\hat\partial_i\ln \mathcal{B}
  ,\quad 
\theta\to \mathcal{B}\theta-\frac{d}{\Omega}\partial_t \mathcal{B}.
\end{equation} 
The Carrollian vorticity $\varpi_{ij}$ \eqref{carconcomderf} and the Carrollian 
shear $\xi_{ij}$ \eqref{dgammaCartime} are Weyl-covariant of weight $-1$. 

The Weyl--Carroll space and time covariant derivatives are metric-compatible.
For a scalar function  $\Phi$ and  a vector $V^l$ of  weight $w$, we find:
\begin{eqnarray}
\label{CWs-Phi}
&&\hat{\mathscr{D}}_j \Phi=\hat\partial_j \Phi +w \varphi_j \Phi,\\
&&\hat{\mathscr{D}}_j V^l=\hat\nabla_j V^l +(w-1) \varphi_j V^l +\varphi^l V_j -\delta^l_j V^i\varphi_i.
\end{eqnarray}
The weights are not altered by the  spatial derivative and $\hat{\mathscr{D}}_j a_{kl}=0$.
One also defines 
\begin{eqnarray}
\label{CWtimecovdersc}
&&\frac{1}{\Omega}\hat{\mathscr{D}}_t \Phi=\frac{1}{\Omega}\hat D_t \Phi +\frac{w}{d} \theta \Phi=
\frac{1}{\Omega}\partial_t \Phi +\frac{w}{d} \theta \Phi,\\
&&\frac{1}{\Omega}\hat{\mathscr{D}}_t V^l=\frac{1}{\Omega}\hat D_t V^l +\frac{w-1}{d} \theta V^l
=\frac{1}{\Omega}\partial_t V^l +\frac{w}{d} \theta V^l+\xi^{l}_{\hphantom{l}i} V^i ,
\label{CWtimecovdervecform}
\end{eqnarray}
both are of weight $w+1$. Furthermore $\hat{\mathscr{D}}_t a_{kl}=0$, using Leibniz rule.

We finally obtain
\begin{eqnarray}
\label{CWcontor}
&&\left[\hat{\mathscr{D}}_i,\hat{\mathscr{D}}_j\right]\Phi=
\frac{2}{\Omega}\varpi_{ij}\hat{\mathscr{D}}_t\Phi
+w \Omega_{ij} 
\Phi,
\\
\label{CWcurvten}
&&\left[\hat{\mathscr{D}}_k,\hat{\mathscr{D}}_l\right]V^i=
\left( \hat{\mathscr{R}}^i_{\hphantom{i}jkl} - 2
\xi^{i}_{\hphantom{i}j}
\varpi_{kl} 
\right)
V^j+
\varpi_{kl}\frac{2}{\Omega}\hat{\mathscr{D}}_t V^i
+w \Omega_{kl}
V^i,
\end{eqnarray}
where  
\begin{eqnarray}
\label{CWRiemann}
\hat{\mathscr{R}}^i_{\hphantom{i}jkl} &=&\hat r^i_{\hphantom{i}jkl}
-\delta^i_j\varphi_{kl}
-a_{jk} \hat{\nabla}_l \varphi^i
+a_{jl} \hat{\nabla}_k \varphi^i 
+\delta^i_k \hat{\nabla}_l \varphi_j 
-\delta^i_l \hat{\nabla}_k \varphi_j 
\nonumber
\\ 
&&+\varphi^i\left(\varphi_k a_{jl}-\varphi_l a_{jk}\right)
-\left(\delta^i_k a_{jl}-\delta^i_l a_{jk}\right)\varphi_m\varphi^m+
\left(\delta^i_k \varphi_l-\delta^i_l \varphi_k\right)\varphi_j,\\
\label{CWOme}
 \Omega_{ij} &=&\varphi_{ij}-\frac{2}{d}\varpi_{ij} \theta,
 \end{eqnarray}
 and $\varphi_{ij}=\hat\partial_i \varphi_j - \hat\partial_j \varphi_i $, are weight-$0$ Weyl-covariant tensors. Tracing them we obtain:
\begin{equation}
\label{CWricci-scalar}
\hat{\mathscr{R}}_{ij}=\hat{\mathscr{R}}^k_{\hphantom{k}ikj}
,\quad \hat{\mathscr{R}}=a^{ij}\hat{\mathscr{R}}_{ij}
\end{equation}
with
\begin{equation}
\label{CWscalar}
\hat{\mathscr{R}}=\hat r +(d-1)\left(2\hat \nabla_i\varphi^i-(d-2)\varphi_i\varphi^i\right),
\end{equation}
of weights zero and $2$. The Weyl-covariant Carroll--Ricci tensor is not symmetric, $\hat{\mathscr{R}}_{[ij]}=-\frac{d}{2} \Omega_{ij}$, and a weight-$1$  curvature form also appears with
\begin{equation}
\left[\frac{1}{\Omega}\hat{\mathscr{D}}_{t},\hat{\mathscr{D}}_i\right]\Phi= w \hat{\mathscr{R}}_{i}\Phi-
\xi^{j}_{\hphantom{j}i}\hat{\mathscr{D}}_j^{\vphantom{j}} \Phi,
\label{CWrsc}
\end{equation}
where
\begin{equation}
\hat{\mathscr{R}}_{i}=
\frac{1}{\Omega} \partial_{t}\varphi_i-\frac{1}{d}\left(\hat \partial_i+\varphi_i\right)\theta.
\label{CWRvec}
\end{equation}

\section{Conformal Carrollian dynamics and charges} \label{conscar}

A complete account on the subject of dynamics and charges with the present conventions is available in Refs. \cite{CMPPS1, BigFluid}. We summarize here the necessary items, in particular regarding the Weyl-covariant side, which is relevant on the holographic boundaries. 

The basics are encoded into four Carrollian momenta, replacing the relativistic energy--momentum tensor, which are obtained by varying some (effective) action with respect to $a_{ij}$, $b_i$ and $\Omega$ (the fourth momentum is not necessarily obtained in this way -- for details see \cite{BigFluid}).  These are the energy--stress tensor $\Pi^{ij}$, the energy flux $\Pi^i$, the energy  density $\Pi$ as well as the momentum $P^i$, of conformal weights $d+3$, $d+2$, $d+1$ and $d+2$. Extra momenta can also emerge as more degrees of freedom may be present. This phenomenon occurs when studying the small-$c$ limit of a relativistic energy--momentum tensor and the corresponding conservation equations. Keeping things rather minimal with only $\tilde\Pi^{ij}$ the equations read:\footnote{\label{bbreaking}Using the language of fluids, $\Pi$ appears as the zero-$c$ limit of the relativistic energy density, $\Pi^i$ and $P^i$ are the orders one and $c^2$ of the relativistic heat current, whereas $\tilde\Pi^{ij}$ and $\Pi^{ij}$ are the orders $\nicefrac{1}{c^2}$ and one of the relativistic stress. A non-vanishing Carrollian energy flux $\Pi^i$ breaks local Carroll-boost invariance (see e.g. \cite{Baiguera:2022lsw}) and makes its dual variable i.e. the Ehresmann connection $\pmb{b}=b_i \text{d}x^i$ dynamical. This is neither a surprise
nor a caveat. On the one hand, Carrollian dynamics, i.e. dynamics on geometries equipped with a degenerate metric, is often reached as a vanishing-$c$ limit of relativistic dynamics and naturally breaks local Carroll boosts, even when the original relativistic theory is Lorentz-boost invariant. Indeed, invariance under local Lorentz boosts sets symmetry constraints on the components of the relativistic energy--momentum tensor, but not on their behaviour with respect to $c^2$, leaving the possibility of persisting energy flux $\Pi^i$  and ``over-stress''  $\tilde\Pi^{ij}$ related through Eq.~\eqref{carHcon}. A similar phenomenon occurs in Galilean theories, defined on spacetimes with a degenerate cometric, where the Galilean momentum is possibly responsible for the breaking of local Galilean-boost invariance. On the other hand, it is fortunate that this happens in the present instance (one of the very few known applications of Carrollian dynamics), when passing from the relativistic boundary of asymptotically anti-de Sitter spacetimes to the Carrollian boundary of their asymptotically flat relatives, as the Carrollian energy flux accounts for non-conservation properties resulting from bulk gravitational radiation, whereas the Ehresmann connection is part of the Ricci-flat solution space.}
\begin{eqnarray}
 \frac{1}{\Omega}\hat{\mathscr{D}}_t\Pi
+\hat{\mathscr{D}}_i \Pi^{i}
+\Pi^{ij}\xi_{ij}&=& 0 ,
  \label{carEbiscon} 
  \\
  \tilde\Pi^{ij}\xi_{ij} &=& 0 ,
  \label{carFcon}
  \\
\hat{\mathscr{D}}_i \Pi^{i}_{\hphantom{i}j}+2\Pi^{i}\varpi_{ij}+ \left(\frac{1}{\Omega}\hat{\mathscr{D}}_t \delta^i_j +\xi^{i}_{\hphantom{i}j}\right) P_i&=&0 ,
  \label{carGcon} 
  \\
\hat{\mathscr{D}}_i \tilde \Pi^{i}_{\hphantom{i}j}+ \left(\frac{1}{\Omega}\hat{\mathscr{D}}_t \delta^i_j +\xi^{i}_{\hphantom{i}j}\right) \Pi_i&=& 0
  \label{carHcon}
 \end{eqnarray}
with 
\begin{equation}
\label{car-conf-cond-more}
\tilde\Pi_i^{\hphantom{i}i}=0,\quad
\Pi_i^{\hphantom{i}i}=\Pi,
\end{equation} 
as a consequence of the assumed Weyl invariance.

Equations \eqref{carEbiscon}, \eqref{carFcon}, \eqref{carGcon} and \eqref{carHcon} are the Carrollian emanation of the relativistic conservation equation $\nabla_\mu T^{\mu\nu}=0$. As for the relativistic instance, conformal isometries lead to conserved currents and conserved charges. Let $\upxi$ be a $d+1$-dimensional vector 
\begin{equation}   
\label{carkil}
\upxi=\xi^t \partial_t +\xi^i \partial_i= \left(\xi^t -\xi^i\frac{b_i}{\Omega}\right) \partial_t
+ \xi^i \left(\partial_i+\frac{b_i}{\Omega}\partial_t\right)=
\xi^{\hat t}\frac{1}{\Omega} \partial_t+\xi^i \hat \partial_i 
 \end{equation}
restricted to $\xi^i=\xi^i(\mathbf{x})$, generator of a one-dimensional group of Carrollian diffeomorphisms on $\mathscr{M}= \mathbb{R} \times \mathscr{S}$. Its action on the elementary geometric data \eqref{cardegmet}, \eqref{kert} and \eqref{kertdual} is as follows:\footnote{The Lie derivative along $\upxi=\xi^{\hat t}\frac{1}{\Omega} \partial_t+\xi^i \hat \partial_i $ of a general Carrollian tensor reads: 
\begin{eqnarray}   
\mathscr{L}_\upxi S_{i\ldots}^{\hphantom{i\ldots}j\ldots}&=&
\left(
\xi^{\hat t}\frac{1}{\Omega} \partial_t+\xi^k \hat \partial_k\right)
S_{i\ldots}^{\hphantom{i\ldots}j\ldots}+
S_{k\ldots}^{\hphantom{k\ldots}j\ldots} \hat \partial_i \xi^k+\cdots-S_{i\ldots}^{\hphantom{i\ldots}k\ldots} \hat \partial_k \xi^j-\cdots
\nonumber\\
&=&
\left(
\xi^{\hat t}\frac{1}{\Omega} \hat D_t+\xi^k \hat \nabla_k\right)
S_{i\ldots}^{\hphantom{i\ldots}j\ldots}+
S_{k\ldots}^{\hphantom{k\ldots}j\ldots} \left(\hat \nabla_i \xi^k + \xi^{\hat t} \hat \gamma^k_{\hphantom{k}i}\right)+\cdots-S_{i\ldots}^{\hphantom{i\ldots}k\ldots} \left(\hat \nabla_k \xi^j+ \xi^{\hat t} \hat \gamma_k^{\hphantom{k}j}\right)-\cdots
.\nonumber
 \end{eqnarray} 
}
\begin{eqnarray}   
\label{Liedaijcar}
\mathscr{L}_\upxi a_{ij}&=& 2\hat\nabla_{(i}\xi^ka_{j)k}+2\xi^{\hat t}  \hat \gamma_{ij}
,\\
\label{Liedcfoocar}
\mathscr{L}_\upxi \upupsilon&=&
\mu
\upupsilon
,\\
\label{Liedcehrecar}
\mathscr{L}_\upxi \upmu&=&-\mu  \upmu +\upnu
\end{eqnarray}
with $\upnu=\nu_i \text{d}x^i$ and 
\begin{equation}
\begin{array}{rcl}
\mu(t,\mathbf{x})&=& -\left(\frac{1}{\Omega} \partial_t
\xi^{\hat t }+
\varphi_i \xi^i \right),
\\
\nu_i(t,\mathbf{x})&=&-\left(\hat\partial_i-\varphi_i\right)\xi^{\hat t }+
2\xi^j \varpi_{ji}.
\label{munui}
\end{array} 
\end{equation} 
Due to the degeneration of the metric on $\mathscr{M}$, the variation of the field of observers $\upupsilon$
is not identical to that of the  clock form $\upmu$. 

Isometries are generated by  Killing fields of the Carrollian type \eqref{carkil}, required to obey  \cite{Duval:2014uva, Duval:2014lpa, Ciambelli:2019lap, Duval:2014uoa}:
\begin{equation}
\label{carkill}
\mathscr{L}_\upxi a_{ij}=0, \quad\mathscr{L}_\upxi \upupsilon=0.
\end{equation}
i.e.
\begin{eqnarray}   
\label{Carolisoa}
\hat\nabla_{(i}\xi^ka_{j)k}+\xi^{\hat t}  \hat \gamma_{ij}&=&0,
\\
\label{CarolisoOm}
\frac{1}{\Omega} \partial_t
\xi^{\hat t }+
\varphi_i \xi^i&=&0.
\end{eqnarray}
The clock form is not required to be invariant.
Carrollian conformal Killing fields  must satisfy 
\begin{equation}
\label{PRkillconf}
\mathscr{L}_\upxi a_{ij}=\lambda a_{ij}
\end{equation}
with 
\begin{equation}
\label{carkilleqconf}
\lambda(t,\mathbf{x}) =\frac{2}{d}
\left(\hat  \nabla_{i}\xi^{i}
+\theta \xi^{\hat t}  
\right).
 \end{equation}
This set of partial differential equations is insufficient for defining conformal Killing fields. One usually imposes to tune $\mu$ versus $\lambda$ (see \cite{Duval:2014uva, Duval:2014lpa, Duval:2014uoa} for a detailed presentation)
 so that the scaling of the metric be twice that of the field of observers: 
\begin{equation}
\label{extra-conf-cond}
2\mu + \lambda=2\left(\frac{1}{d}
\hat{\mathscr{D}}_i\xi^i -\frac{1}{\Omega} \hat{\mathscr{D}}_t
\xi^{\hat t }
\right)
=0
 \end{equation}
(the conformal weight of $\xi^{\hat t} $ is $-1$, that of $\xi^i$ is zero). Again, the clock form is not involved. If one demands the latter be invariant under the action of a Killing field, or aligned with itself under the action of a conformal Killing, which in both cases amounts to setting 
\begin{equation}
\label{super-K}
\nu_i
\equiv-\hat{\mathscr{D}}_i\xi^{\hat t }+
2\xi^j \varpi_{ji} =0
 \end{equation}
(this is a conformal rewriting of $\nu_i$ given in \eqref{munui}), then the corresponding (conformal) isometry generator will be referred to as \emph{(conformal) strong Killing vector field}.\footnote{Carroll boosts, which are the archetype isometries of flat Carrollian spacetimes, \emph{are not} generated by strong Killings~\cite{BigFluid}.}

Transformations generated by ordinary Carrollian Killing fields leave invariant geometric markers that are built over the metric $a_{ij}$ and $\Omega$, such as the Carrollian expansion $\theta$ or the shear $\xi_{ij}$, encoded in $\hat \gamma_{ij}$, see \eqref{dgammaCartime}. The Carrollian vorticity and acceleration given in 
\eqref{dualcarconcomderf} are not left invariant, however, since
\begin{equation}
\label{Liedcardmu}
\mathscr{L}_\upxi \text{d}\upmu=
\text{d}\upnu,
 \end{equation}
unless the Carrollian Killing field is strong (vanishing $\upnu$). Likewise, curvature invariance does also require the strong condition. This applies in particular to the Carrollian Cotton tensor discussed in $d=2$ (see App. \ref{carcot3}). 
 
On a Carroll manifold a current has a scalar component $\kappa$ as well as a Carrollian-vector set of components $K^i$. The divergence takes the form (see \cite{CM1,BigFluid, Rivera-Betancour:2022lkc})
 \begin{equation}
\mathcal{K}= \left(
\frac{1}{\Omega}\partial_t+
\theta \right)\kappa
+\left(\hat \nabla_i +\varphi_i\right)K^i.
 \label{kilcarcon}
 \end{equation}
The charge associated with the current $(\kappa, \pmb{K})$ is an integral at fixed $t$ over the basis $\mathscr{S}$   
\begin{equation}
Q_K=\int_{\mathscr{S}}\text{d}^{d}x \sqrt{a}\left(\kappa+b_iK^i \right),
  \label{carconch}
 \end{equation}
and obeys the following time evolution:
  \begin{equation}
\frac{\text{d}Q_K}{\text{d}t}=\int_{\mathscr{S}}\text{d}^{d}x \sqrt{a}\Omega \mathcal{K}
- \int_{\partial\mathscr{S}} \ast\pmb{K} \, \Omega.
\label{carconchdt}
 \end{equation}
 The last term is of boundary type with $\ast\pmb{K}$ the $\mathscr{S}$-Hodge dual of $K_i\text{d}x^i$.  
 Generally, one can ignore it owing to adequate fall-off or boundary conditions on the fields. 
 
Suppose that $\upxi$ is the generator \eqref{carkil}  of a Carrollian diffeomorphism. It can be used to create two currents out\footnote{We stress here that if more momenta were present, more currents would be available.} of $\Pi^{ij}$, $\tilde \Pi^{ij}$, $\Pi^{i}$,  $P^i$
and
$\Pi$   \cite{CM1, BigFluid}:
\begin{equation}
\label{carkappaK}
 \begin{cases}
\kappa= \xi^{i} P_i-\xi^{\hat t}  \Pi
\\
\tilde\kappa=\xi^{i} \Pi_i
\\
K^i=\xi^{j}\Pi_{j}^{\hphantom{j}i}-\xi^{\hat t} \Pi^i
\\
\tilde K^i =\xi^{j}\tilde \Pi_{j}^{\hphantom{j}i}
,
\end{cases}
\end{equation}
If $\upxi$ is a (conformal) Carrollian Killing field, and assuming all momenta on-shell i.e. Eqs.
 \eqref{carEbiscon}, \eqref{carFcon}, \eqref{carGcon} and \eqref{carHcon} (with \eqref{car-conf-cond-more} satisfied in the conformal instance), one finds the following Carrollian divergences (the conformal weights of $\kappa$ and $\tilde \kappa$ are $d$, those of $K^i$ and $\tilde K^i$, $d+1$, and $-1$ for $\nu_i$):\footnote{Notice that these are the specific conformal weights ensuring  the  Carroll divergence in \eqref{kilcarcon} be identical to the Weyl--Carroll divergence. For $\left\{\kappa, K^i\right\}$ of general weights $(w,w+1)$ we find instead $ \left(
\frac{1}{\Omega}\partial_t+
\theta \right)\kappa
+\left(\hat \nabla_i +\varphi_i\right)K^i = 
  \frac{1}{\Omega}\hat{\mathscr{D}}_t\kappa
+\hat{\mathscr{D}}_i  K^i
+ (d-w)\left(\frac{\theta}{d}\kappa+\varphi_i K^i\right)
$.}
\begin{equation}
\label{carKN}
 \begin{cases}
\tilde{\mathcal{K}}=
 \frac{1}{\Omega}\hat{\mathscr{D}}_t\tilde\kappa
+\hat{\mathscr{D}}_i \tilde K^i
=
0
 \\ \mathcal{K}=
 \frac{1}{\Omega}\hat{\mathscr{D}}_t\kappa
+\hat{\mathscr{D}}_i  K^i
=\Pi^i\nu_i.
\end{cases}
\end{equation}
Two charges can be defined following \eqref{carconch}: 
$Q_{\tilde K}$ and $Q_K$. The former is conserved, whereas the latter isn't for generic isometries unless the field configuration has vanishing energy flux $\Pi^i$, i.e. if local Carroll-boost invariance is unbroken. The breaking of local Carroll-boost invariance hence appears as the trigger of non-conservation laws. This peculiarity was risen in \cite{CM1,BigFluid} and further illustrated with concrete field realizations in \cite{Rivera-Betancour:2022lkc}. In four-dimensional Ricci-flat spacetimes, this boundary non-conservation is the consequence of bulk gravitational radiation, as mentioned previously in footnote \ref{bbreaking}. Observe nevertheless that irrespective of the energy flux $\Pi^i$, the (conformal) strong Killings introduced earlier do lead to full conservation properties as a consequence of \eqref{super-K}.

\section{Three dimensions and the Carrollian Cotton tensor} \label{carcot3}

Three-dimensional boundaries ($d=2$) outline the framework of the Ehlers and Geroch investigation pursued in the main part of this article. Three dimensions have two remarkable properties.  At the first place, if the geometric Carrollian shear $\xi^{ij}$
defined in \eqref{dgammaCartime} vanishes, which occurs for the Carrollian boundaries of Ricci-flat spacetimes as a consequence of Einstein's equations (see Sec. \ref{brrm}), the Carrollian conformal  isometry group is infinite-dimensional: $\text{BMS}_{4}\equiv\mathfrak{ccarr}(3)\equiv \mathfrak{so}(3,1)\ltimes \text{supertranslations}$ \cite{CMPPS2, Ciambelli:2019lap}. This potentially generates \emph{infinite towers of charges}, possibly conserved. 

Secondly, three-dimensional Carrollian spacetimes possess a \emph{Carrollian Cotton tensor obeying conservation dynamics}. It appears as a set of Carrollian scalars, vectors and tensors emerging in the small-$c$ expansion of the relativistic Cotton $C_{\mu\nu}$, which is symmetric, traceless, divergence-free and Weyl-covariant with weight $1$. Reference \cite{CMPPS2} provides a complete account of the Carrollian descendants as they emerge from the pseudo-Riemannian Cotton tensor, in the absence of geometric Carrollian shear. Here we will circumscribe our exhibition to the basic output. 

For $d=2$, the  $\mathscr{S}$-Hodge duality is induced by\footnote{We use here the conventions of Ref. \cite{CMPPS2}, namely $\epsilon_{12}=-1$, convenient when using complex coordinates $\{\zeta, \bar \zeta\}$. Notice that $\eta^{il}\eta_{jl}=\delta^i_j$ and  $\eta^{ij}\eta_{ij}=2$. }  $\eta_{ij} = \sqrt{a}\epsilon_{ij}$.  This duality is involutive on Carrollian vectors as well as on two-index symmetric and traceless Carrollian tensors:
\begin{equation}
\label{hodgeast}
\ast\! V_i=\eta^l_{\hphantom{l}i}V_l, \quad \ast W_{ij}=\eta^l_{\hphantom{l}i}W_{lj}.
\end{equation}
This fully antisymmetric form can be used to recast some of the expressions introduced in App. \ref{carman}. The Carroll--Ricci tensor \eqref{carricci-scalar} is decomposed as 
\begin{equation}
\label{carricci-expand}
\hat r_{ij}=\hat s_{ij}+\hat K a_{ij}+\hat A \eta_{ij}
\end{equation}
with
\begin{equation}
\label{scalar}
\hat s_{ij}=2\ast\!  \varpi \ast\!  \xi_{ij},\quad \hat K=\frac{1}{2} a^{ij}\hat r_{ij}=\frac{1}{2} \hat r, \quad \hat A=\frac{1}{2} \eta^{ij}\hat r_{ij}=\ast\varpi\theta,\quad \ast\varpi=\frac{1}{2} \eta^{ij} \varpi_{ij}.
\end{equation}
Similarly 
\begin{equation}
\label{CWricci-dec}
\hat{\mathscr{R}}_{ij}=\hat s_{ij}+\hat{\mathscr{K}} a_{ij}+\hat{\mathscr{A}} \eta_{ij},
\end{equation}
where we have introduced two weight-$2$ Weyl-covariant scalar Gauss--Carroll curvatures:
\begin{equation}
\label{CWscalar}
\hat{\mathscr{K}}=\frac{1}{2}a^{ij}\hat{\mathscr{R}}_{ij}=\hat{K}+ \hat{\nabla}_k \varphi^k
,\quad \hat{\mathscr{A}}=\frac{1}{2}\eta^{ij}\hat{\mathscr{R}}_{ij}=  \hat{A}- \ast \varphi,
\end{equation}
and  $\ast\varphi=\frac{1}{2} \eta^{ij} \varphi_{ij} $. These obey Carroll-Bianchi identities:
\begin{eqnarray}
 \frac{2}{\Omega}\hat{\mathscr{D}}_t \ast\! \varpi +\hat{\mathscr{A}}&=&0
\label{Carroll-Bianchi1}
,\\
 \frac{1}{\Omega}\hat{\mathscr{D}}_t \hat{\mathscr{K}}
-a^{ij}\hat{\mathscr{D}}_i  \hat{\mathscr{R}}_{j} -  \hat{\mathscr{D}}_i \hat{\mathscr{D}}_j \xi^{ij} &=& 0 ,
\label{Carroll-Bianchi2}
\\
\frac{1}{\Omega}\hat{\mathscr{D}}_t \hat{\mathscr{A}}+ \eta^{ij}\hat{\mathscr{D}}_i  \hat{\mathscr{R}}_{j} &=&0
\label{Carroll-Bianchi3}
.
\end{eqnarray}

Thanks to the identities \eqref{Carroll-Bianchi2} and \eqref{Carroll-Bianchi3}, the couples $\left\{\hat{\mathscr{K}},-\hat{\mathscr{R}}^{i}- \hat{\mathscr{D}}_j \xi^{ij}\right\}$ and $\left\{\hat{\mathscr{A}},-\ast\! \hat{\mathscr{R}}^{i} \right\}$ of weights $(2,3)$ allow to define \emph{electric and magnetic curvature charges} as in Eqs. \eqref{kilcarcon} and \eqref{carconch}:
\begin{equation}
Q_{\text{ec}}=\int_{\mathscr{S}}\text{d}^{2}x \sqrt{a}\left(\hat{\mathscr{K}}-b_i \left(\hat{\mathscr{R}}^{i}+\hat{\mathscr{D}}_j \xi^{ij}\right)\right),\quad
Q_{\text{mc}}=\int_{\mathscr{S}}\text{d}^{2}x \sqrt{a}\left(\hat{\mathscr{A}}-b_i \ast\!\hat{\mathscr{R}}^{i}\right).
  \label{elmgcarconch}
 \end{equation}
Following \eqref{carconchdt}, we find 
  \begin{equation}
\frac{\text{d}Q_{\text{ec}}}{\text{d}t}=  \int_{\partial\mathscr{S}} \ast \left(\hat{\pmb{\mathscr{R}}}+\hat{\pmb{\mathscr{D}}}\cdot\pmb{\xi}\right) \Omega, \quad
\frac{\text{d}Q_{\text{mc}}}{\text{d}t}= - \int_{\partial\mathscr{S}} \hat{\pmb{\mathscr{R}}}\,  \Omega.
\label{elmgcarconchdt}
 \end{equation}
Upon regular behaviour, the boundary terms vanish and the curvature charges are both conserved. 

Besides the various curvature tensors, which are second derivatives of the metric and the Ehresmann connection, one defines third-derivative tensors, the descendants of the relativistic Cotton tensor. We will here limit our presentation to the instance $\xi_{ij}=0$, which is the appropriate framework when solving Einstein's equations in the bulk. This reduces the number of tensors to five, a weight-$3$ scalar, two weight-$2$ forms and two weight-$1$ two-index symmetric and traceless tensors:
\begin{eqnarray}
\label{c-Carrol}
c&=&\left(\hat{\mathscr{D}}_l\hat{\mathscr{D}}^l+2\hat{\mathscr{K}}
\right)\ast\!\varpi,
\\
\label{chi-f-Carrol}
\chi_j&=&\frac{1}{2}\eta^l_{\hphantom{l}j}\hat{\mathscr{D}}_l\hat{\mathscr{K}}+ \frac{1}{2} \hat{\mathscr{D}}_j\hat{\mathscr{A}}-2\ast\!\varpi\hat{\mathscr{R}}_j,
\\
\label{psi-f-Carrol}
\psi_j&=&3\eta^l_{\hphantom{l}j}\hat{\mathscr{D}}_l\ast\!\varpi^2,
\\
\label{X-2-Carrol}
X_{ij}&=&\frac{1}{2}\eta^l_{\hphantom{l}j}\hat{\mathscr{D}}_l
\hat{\mathscr{R}}_i+
\frac{1}{2} \eta^l_{\hphantom{l}i}\hat{\mathscr{D}}_j
\hat{\mathscr{R}}_l,
\\
\label{Psi-2-Carrol}
\Psi_{ij}&=&\hat{\mathscr{D}}_i \hat{\mathscr{D}}_j\ast\!\varpi -\frac{1}{2}a_{ij} \hat{\mathscr{D}}_l \hat{\mathscr{D}}^l \ast\!\varpi -\eta_{ij} \frac{1}{\Omega}  \hat{\mathscr{D}}_t\ast\!\varpi^2.
\end{eqnarray}  
As a consequence of the relativistic conservation of the Cotton tensor, its Carrollian descendants obey Eqs.  \eqref{carEbiscon}, \eqref{carFcon},\footnote{Equation \eqref{carFcon}, is trivially satisfied due to the vanishing of $\xi_{ij}$. If $\xi_{ij}\neq 0$, extra Cotton Carrollian descendants are available, and the conservation dynamics is encoded in more momenta and equations -- in particular  \eqref{carFcon} is modified. } \eqref{carGcon} and \eqref{carHcon}  with
\begin{equation}
\label{mom-Cot}
\Pi_{\text{Cot}}=c,\quad \Pi_{\text{Cot}}^i=\chi^i,\quad P_{\text{Cot}}^i=\psi^i,\quad \tilde\Pi_{\text{Cot}}^{ij}=-X^{ij},\quad \Pi_{\text{Cot}}^{ij}=\frac{c}{2}a^{ij}-\Psi^{ij},
\end{equation}
which read
\begin{eqnarray}
\frac{1}{\Omega}\hat{\mathscr{D}}_t c+\hat{\mathscr{D}}_i \chi^{i}
&=&0,
 \label{carEcot} 
\\
\frac{1}{2}\hat{\mathscr{D}}_j c
+2\chi^{i}\varpi_{ij}
+ \frac{1}{\Omega}\hat{\mathscr{D}}_t \psi_j
- \hat{\mathscr{D}}_i \Psi^i_{\hphantom{i}j}
&=& 0,
  \label{carGcot}\\
\frac{1}{\Omega}\hat{\mathscr{D}}_t 
\chi_{j}- \hat{\mathscr{D}}_i X^{i}_{\hphantom{i}j}
&=&0.
 \label{carHcot} 
\end{eqnarray}

When the geometric Carrollian shear vanishes, the time dependence in the metric is factorized as $a_{ij}(t,\mathbf{x})=\text{e}^{2\sigma(t,\mathbf{x})}\bar a_{ij}(\mathbf{x})$. One then shows \cite{CMPPS2, Ciambelli:2019lap} that
the Carrollian conformal isometry group is the semi-direct product of the conformal group of $\bar a_{ij}(\mathbf{x})$ with the infinite-dimensional supertranslation group. The former is generated by $Y^i(\mathbf{x})$, the latter by $T(\mathbf{x})$, and the Carrollian conformal Killing fields read:
\begin{equation}
\label{Xclock}
\upxi_{{T},Y}=\left({T}(\mathbf{x})
-Y^i(\mathbf{x})\hat\partial_iC(t,\mathbf{x})+\frac{1}{2}C(t,\mathbf{x})\bar{\nabla}_iY^i(\mathbf{x})\right)\frac{\text{e}^{\sigma(t,\mathbf{x})}}{\Omega}\partial_t+Y^i(\mathbf{x})\hat\partial_i
\end{equation}
with
\begin{equation}
\label{clock}
C(t,\mathbf{x})\equiv \int^t \text{d}{\tau}\; \text{e}^{-\sigma(\tau,\mathbf{x})}\Omega\left({\tau},\mathbf{x}\right).
\end{equation}
This result is valid in any dimension. At $d=2$, $\bar a_{ij}(\mathbf{x})$ is conformally flat and $Y^i(\mathbf{x})$ generate  $ \mathfrak{so}(3,1)$.\footnote{The $ \mathfrak{so}(3,1)$ factor can also be promoted to superrotations (double Virasoro) if we give up the absolute regularity requirement.}

The conservation of the Carrollian Cotton momenta \eqref{mom-Cot} makes it possible to define two infinite towers of Carrollian Cotton charges $Q_{\text{Cot}\, T,Y}$ and $\tilde Q_{\text{Cot}\, T,Y}$ following \eqref{carconch}, based on the Carrollian Cotton currents $\kappa_{\text{Cot}}$, $K^i_{\text{Cot}}$, $\tilde \kappa_{\text{Cot}}$ and  $\tilde K^i_{\text{Cot}}$
 (see \eqref{carkappaK-cot-res}). According to \eqref{carKN}, the latter are always conserved,\footnote{The conformal Killing fields \eqref{Xclock}, \eqref{clock} depend \emph{explicitly} on time. Inside the charges they define, when conserved, this time dependence is confined, on-shell, in a boundary term, and hence drops -- see concrete examples in \cite{Rivera-Betancour:2022lkc}.} whereas the former are only if 
  $\chi^i\nu_i=-\chi^i\left(\hat{\mathscr{D}}_i\xi^{\hat t }-
2\xi^j \varpi_{ji}
\right)=0
$. This occurs for special geometries ($\chi^i=0$) or for the subset of strong Carrollian conformal Killing fields ($\nu_i=0$) .

In $d=2$, it is convenient to use complex spatial coordinates $\zeta$ and $\bar \zeta$. With the permission of the authors of \cite{CMPPS2}, we reproduce here the appendix of that reference, summarizing the useful formulas in this coordinate system.
Using Carrollian diffeomorphisms \eqref{cardifs}, the metric \eqref{cardegmet} of the Carrollian geometry on the two-dimensional surface
$\mathscr{S}$ can be recast in conformally flat form,
\begin{equation}
\label{CF}
\text{d}\ell^2=\frac{2}{P^2}\text{d}\zeta\text{d}\bar\zeta
\end{equation}
with $P=P(t,\zeta,\bar \zeta)$ a real function, under the necessary and sufficient condition that the Carrollian shear $\xi_{ij}$ displayed in \eqref{dgammaCartime} vanishes. We will here assume that this holds and present a number of useful formulas for Carrollian and conformal Carrollian geometry.  
  These geometries carry two further pieces of data: $\Omega(t,\zeta,\bar \zeta)$ and  
\begin{equation}
\label{frame}
\pmb{b}=b_\zeta(t,\zeta, \bar \zeta)\, \text{d}\zeta+b_{\bar\zeta}(t,\zeta, \bar \zeta)\, \text{d}\bar\zeta
\end{equation}
with $b_{\bar\zeta}(t,\zeta, \bar \zeta)=\bar b_\zeta(t,\zeta, \bar \zeta)$. 
Our choice of orientation is inherited from the one adopted for the relativistic boundary (see footnote \ref{orient}) with  $a_{\zeta\bar\zeta}=\nicefrac{1}{P^2}$ is\footnote{This amounts to setting $\sqrt{a}=\nicefrac{\text{i}}{P^2}$ in coordinate frame and $\epsilon_{\zeta\bar\zeta}=-1$. The volume form reads $\text{d}^2x \sqrt{a}=\frac{\text{d}\zeta\wedge \text{d}\bar\zeta}{\text{i}P^2}$.}
\begin{equation}
\label{orientCF}
\eta_{\zeta\bar\zeta}=-\frac{\text{i}}{P^2}.
\end{equation}

The first-derivative Carrollian tensors are the acceleration \eqref{carconcomderf}, the expansion \eqref{dgammaCartime} and 
the scalar vorticity \eqref{carconcomderf}, \eqref{scalar}:
\begin{eqnarray}
&\varphi_{\zeta}
=\partial_t \dfrac{b_{\zeta}}{\Omega}+\hat\partial_{\zeta} \ln \Omega,\quad \varphi_{\bar\zeta}
=\partial_t \dfrac{b_{\bar\zeta}}{\Omega}+\hat\partial_{\bar\zeta} \ln \Omega,&
\label{acchol}
\\
&\theta =-\dfrac{2}{\Omega} \partial_t \ln P,
\quad 
\ast\varpi=\dfrac{\text{i}\Omega P^2}{2}\left(
\hat\partial_{\zeta}\dfrac{b_{\bar\zeta}}{\Omega}-\hat\partial_{\bar\zeta} \dfrac{ b_{\zeta}}{\Omega}
\right)&
\label{thetstarvarpihol}
\end{eqnarray}
with
\begin{equation}
\hat\partial_{\zeta} = \partial_\zeta+\frac{b_{\zeta}}{\Omega}\partial_t,\quad
\hat\partial_{\bar\zeta} = \partial_{\bar\zeta}+\frac{b_{\bar\zeta}}{\Omega}\partial_t.
\end{equation}  
Curvature scalars and vector are second-derivative (see \eqref{scalar},  \eqref{carriemanntime}):\footnote{We also quote for completeness (useful \emph{e.g.} in Eq. \eqref{CWscalarhol}):
$$
\hat K =K+ P^2 \left[ 
\partial_{\zeta} \frac{b_{\bar\zeta}}{\Omega}+
\partial_{\bar\zeta} \frac{b_{\zeta}}{\Omega}+
\partial_{t} \frac{b_{\zeta}b_{\bar\zeta}}{\Omega^2}
+2  \frac{b_{\bar\zeta}}{\Omega}\partial_{\zeta}
+2 \frac{b_{\zeta}}{\Omega} \partial_{\bar\zeta}
+2 \frac{b_{\zeta}b_{\bar\zeta}}{\Omega^2}\partial_{t}
 \right]\partial_t\ln P
$$
with $K= 2P^2 \partial_{\bar\zeta} \partial_\zeta\ln P$ 
the ordinary Gaussian curvature of the two-dimensional metric 
 \eqref{CF}.}
\begin{eqnarray}
&\hat K=P^2\left(\hat\partial_{\bar\zeta} 
 \hat\partial_{\zeta}
 + \hat\partial_{\zeta}\hat\partial_{\bar\zeta}
 \right)
\ln P,
\quad
\hat A=
\text{i}P^2\left(\hat\partial_{\bar\zeta} 
 \hat\partial_{\zeta}
 - \hat\partial_{\zeta}\hat\partial_{\bar\zeta}
 \right)
\ln P,&
\label{hatKA}
\\
&
\hat r_\zeta=\dfrac{1}{2}\hat\partial_{\zeta}\left(\dfrac{1}{\Omega}
\partial_t\ln P
\right),\quad
\hat r_{\bar\zeta}=\dfrac{1}{2}\hat\partial_{\bar\zeta}\left(\dfrac{1}{\Omega}
\partial_t\ln P
\right),
&
\label{hatr}
\end{eqnarray}
and we also quote:
\begin{eqnarray}
&
\ast \varphi=\text{i}P^2\left(
\hat\partial_{\zeta}\varphi_{\bar\zeta}-\hat\partial_{\bar\zeta} \varphi_{\zeta}
\right),
&
\\
&\hat\nabla_k\varphi^k= P^2\left[\hat\partial_\zeta \partial_t \frac{b_{\bar \zeta}}{\Omega}
+
\hat\partial_{\bar \zeta} \partial_t \frac{b_\zeta}{\Omega}
+\left(\hat\partial_{\zeta}\hat\partial_{\bar \zeta}+\hat\partial_{\bar \zeta}\hat\partial_\zeta
\right)\ln \Omega
\right].&
\end{eqnarray}

Regarding conformal Carrollian tensors we remind the weight-$2$ curvature scalars \eqref{CWscalar}:
\begin{equation}
\label{CWscalarhol}
\hat{\mathscr{K}}=\hat{K}+ \hat{\nabla}_k \varphi^k
,\quad \hat{\mathscr{A}}=  \hat{A}- \ast \varphi,
\end{equation}
and the  weight-$1$ curvature one-form \eqref{CWRvec}:
\begin{equation}
\hat{\mathscr{R}}_{\zeta}=\frac{1}{\Omega} \partial_{t}\varphi_{\zeta}-\frac{1}{2}\left(\hat \partial_{\zeta}+\varphi_{\zeta}\right)\theta, \quad \hat{\mathscr{R}}_{\bar\zeta}=\frac{1}{\Omega} \partial_{t}\varphi_{\bar\zeta}-\frac{1}{2}\left(\hat \partial_{\bar\zeta}+\varphi_{\bar\zeta}\right)\theta.
\label{CWRvechol}
\end{equation}
The three-derivative Cotton descendants displayed in \eqref{c-Carrol}--\eqref{Psi-2-Carrol} are a scalar
\begin{equation}
\label{c-Carrolhol}
c=\left(\hat{\mathscr{D}}_l\hat{\mathscr{D}}^l+2\hat{\mathscr{K}}
\right)\ast\!\varpi
\end{equation}  
of weight $3$ ($\ast \varpi$ is of weght $1$), two vectors
\begin{eqnarray}
\label{chi-f-Carrolhol}
&\chi_{\zeta}=\frac{\text{i}}{2}\hat{\mathscr{D}}_{\zeta}\hat{\mathscr{K}}+ \frac{1}{2} \hat{\mathscr{D}}_{\zeta}\hat{\mathscr{A}}-2\ast \! \varpi\hat{\mathscr{R}}_{\zeta},
\quad 
\chi_{\bar\zeta}=-\frac{\text{i}}{2}\hat{\mathscr{D}}_{\bar\zeta}\hat{\mathscr{K}}+ \frac{1}{2} \hat{\mathscr{D}}_{\bar\zeta}\hat{\mathscr{A}}-2\ast\!  \varpi\hat{\mathscr{R}}_{\bar\zeta}
,
&
\\
\label{psi-f-Carrolhol}
&\psi_{\zeta}=3\text{i}\hat{\mathscr{D}}_{\zeta}\ast\!  \varpi^2,
\quad \psi_{\bar\zeta}=-3\text{i}\hat{\mathscr{D}}_{\bar\zeta}\ast\!  \varpi^2,&
\end{eqnarray}  
of weight $2$, and two symmetric and traceless tensors
\begin{eqnarray}
\label{X-2-Carrolhol}
&X_{\zeta\zeta}=\text{i}\hat{\mathscr{D}}_{\zeta}
\hat{\mathscr{R}}_{\zeta},\quad
X_{{\bar\zeta}{\bar\zeta}}=-\text{i}\hat{\mathscr{D}}_{{\bar\zeta}}
\hat{\mathscr{R}}_{{\bar\zeta}}
,&
\\
\label{Psi-2-Carrolhol}
&\Psi_{\zeta\zeta}=\hat{\mathscr{D}}_{\zeta} \hat{\mathscr{D}}_{\zeta}\ast\!  \varpi ,
\quad 
\Psi_{{\bar\zeta}{\bar\zeta}}=\hat{\mathscr{D}}_{{\bar\zeta}} \hat{\mathscr{D}}_{{\bar\zeta}}\ast\!  \varpi ,
&
\end{eqnarray}  
of weight $1$. Notice that in holomorphic coordinates a symmetric and traceless tensor $S_{ij}$ has only diagonal entries: $S_{\zeta\bar{\zeta}}=0=S_{\bar{\zeta}\zeta}$. 

We  also remind for convenience some expressions for the determination of Weyl--Carroll covariant derivatives. If $\Phi$ is a weight-$w$ scalar function
\begin{equation}
\label{CWs-Phi-hol}
\hat{\mathscr{D}}_{\zeta} \Phi=\hat\partial_{\zeta} \Phi +w \varphi_{\zeta} \Phi,\quad \hat{\mathscr{D}}_{\bar\zeta} \Phi=\hat\partial_{\bar\zeta} \Phi +w \varphi_{\bar\zeta} \Phi.
\end{equation}
For weight-$w$ form components $V_{\zeta} $ and $V_{\bar\zeta} $ the Weyl--Carroll derivatives read: 
\begin{eqnarray}
&\hat{\mathscr{D}}_{\zeta} V_{\zeta}=\hat\nabla_{\zeta} V_{\zeta} +(w+2) \varphi_{\zeta}V_{\zeta},\quad
\hat{\mathscr{D}}_{\bar\zeta} V_{\bar\zeta}=\hat\nabla_{\bar\zeta} V_{\bar\zeta} +(w+2) \varphi_{\bar\zeta}V_{\bar\zeta},&
\\
&\hat{\mathscr{D}}_{\zeta} V_{\bar\zeta}=\hat\nabla_{\zeta} V_{\bar\zeta} +w \varphi_{\zeta}V_{\bar\zeta},\quad
\hat{\mathscr{D}}_{\bar\zeta} V_{\zeta}=\hat\nabla_{\bar\zeta} V_{\zeta} +w \varphi_{\bar\zeta}V_{\zeta},&
\end{eqnarray}
while the Carrollian covariant derivatives are simply:
\begin{eqnarray}
&\hat\nabla_{\zeta} V_{\zeta} =\dfrac{1}{P^2}\hat\partial_{\zeta}\left(P^2
V_{\zeta}
\right),\quad
\hat\nabla_{\bar\zeta} V_{\bar\zeta} =\dfrac{1}{P^2}\hat\partial_{\bar\zeta}\left(P^2
V_{\bar\zeta}
\right),&
\\
&\hat\nabla_{\zeta} V_{\bar\zeta}= \hat\partial_{\zeta}V_{\bar\zeta},\quad
\hat\nabla_{\bar\zeta} V_{\zeta} =\hat\partial_{\bar\zeta}V_{\zeta}.&
\end{eqnarray}
Finally,
\begin{equation}
\label{weyllapl-scalhol}
\hat{\mathscr{D}}_k\hat{\mathscr{D}}^k \Phi=P^2\left(
 \hat\partial_{\zeta}\hat\partial_{\bar\zeta} \Phi +\hat\partial_{\bar\zeta} \hat\partial_{\zeta} \Phi 
+w \Phi\left( \hat\partial_{\zeta} \varphi_{\bar\zeta} 
+
 \hat\partial_{\bar\zeta} \varphi_\zeta
\right)
+2w\left(\varphi_{\zeta} \hat\partial_{\bar\zeta} \Phi +\varphi_{\bar\zeta} \hat\partial_{\zeta} \Phi 
+w\varphi_{\zeta} \varphi_{\bar\zeta} 
\Phi 
\right)\right).
\end{equation}  

Using complex coordinates, we can recast the conformal Killing vectors of a shear-free Carrollian spacetime $\mathscr{M}$ in  three dimensions, 
given in Eqs. \eqref{Xclock} and \eqref{clock}.
These
are expressed in terms of an arbitrary real function $T( \zeta, \bar \zeta)$, which encodes the \emph{supertranslations}, and the conformal Killing vectors of flat space $\text{d}\bar \ell^2=2\text{d}\zeta\text{d}\bar\zeta$. The latter are of the form 
$Y^\zeta(\zeta)\, \partial_\zeta+Y^{\bar\zeta}(\bar\zeta)\,  \partial_{\bar\zeta}$, reached with any   combination of  $\ell_m+\bar \ell_m$ or $\text{i}\left(\ell_m-\bar \ell_m\right)$, where\footnote{Notice that combining \eqref{hodgeast} and \eqref{orientCF}, we find $\ast\left(\ell_m+\bar \ell_m \right)=-\text{i}\left(\ell_m-\bar \ell_m\right)$. \label{astkil}} 
\begin{equation}
\label{lmlbarm} 
\ell_m=-\zeta^{m+1}\partial_\zeta, \quad  \bar \ell_m = -\bar\zeta^{m+1}\partial_{\bar\zeta},
\end{equation}
obeying the $\text{Witt}\oplus \text{Witt}$ algebra:
\begin{equation}
\label{wittwitt} 
\left[\ell_m, \ell_n
\right]=(m-n)\ell_{m+n}, \quad
\left[\bar\ell_m,\bar \ell_n
\right]=(m-n)\bar\ell_{m+n},
\end{equation}
and referred to as \emph{superrotations}. Usually one restricts to $\mathfrak{so}(3,1)$, generated by $n=0, \pm 1$.
The conformal Killing fields of $\mathscr{M}$ are thus
\begin{equation}
\label{comXclock}
\upxi_{T,Y}
=
\left(T
-\left(Y^\zeta \hat\partial_\zeta+Y^{\bar\zeta} \hat\partial_{\bar\zeta}\right)C+\frac{C}{2}\left(\partial_\zeta Y^\zeta+\partial_{\bar\zeta}Y^{\bar \zeta} \right)\right)\frac{1}{P}\upupsilon+Y^\zeta \hat\partial_\zeta+Y^{\bar\zeta} \hat\partial_{\bar\zeta}
\end{equation}
with
\begin{equation}
\label{comclock}
C(t,\zeta,\bar \zeta)\equiv \int^t \text{d}{\tau}\,  P(\tau, \zeta,\bar \zeta)\, \Omega(\tau, \zeta,\bar \zeta).
\end{equation}
The structure $\mathfrak{so}(3,1)\loplus \text{supertranslations}$ -- or $\left(\text{Witt}\oplus \text{Witt}\right)\loplus \text{supertranslations}$ -- is recovered in
\begin{equation}
\left[ \upxi_{{T}, Y},\upxi_{{T}', Y'}\right] =\upxi_{{M}_{Y}({T}')-{M}_{ Y'}({T}),[ Y, Y']}
\end{equation}
with 
\begin{equation}
\label{MY} 
M_Y(f)=\left(Y^\zeta \hat\partial_\zeta+Y^{\bar\zeta} \hat\partial_{\bar\zeta}\right)f-\frac{f}{2}\left(\partial_\zeta Y^\zeta+\partial_{\bar\zeta}Y^{\bar \zeta} \right). 
\end{equation}


\end{document}